\pdfoutput=1
\RequirePackage{ifpdf}
\ifpdf 
\documentclass[pdftex]{sigma}
\else
\documentclass{sigma}
\fi

\usepackage{epic}
\usepackage{tikz}

\numberwithin{equation}{section}

\newtheorem{Theorem}{Theorem}[section]
\newtheorem*{Theorem*}{Theorem}

 { \theoremstyle{definition}

\newtheorem{Remark}[Theorem]{Remark} }

\newcommand{\wb}[1]{\overline{#1}}

\newcommand{\wh}{\widehat}
\newcommand{\wt}{\widetilde}

\newcommand{\Ld}{\boldsymbol{\Lambda}}
\newcommand{\tLd}{\,^{{\rm t}\!}\boldsymbol{\Lambda}}
\newcommand{\tc}{\,^{{\rm t}\!}\boldsymbol{c}}

\newcommand{\bOm}{\boldsymbol{\Omega}}

\newcommand{\tbme}{\,^{{\rm t}\!}{\boldsymbol{e}}}
\newcommand{\bC}{\boldsymbol{C}}
\newcommand{\bL}{\boldsymbol{L}}
\newcommand{\bM}{\boldsymbol{M}}
\newcommand{\bO}{\boldsymbol{O}}
\newcommand{\bU}{\boldsymbol{U}}

\newcommand{\sg}{\sigma}
\newcommand{\ld}{\lambda}
\newcommand{\oa}{\omega}
\newcommand{\nn}{\nonumber}

\newcommand{\bu}{\boldsymbol{u}}

\newcommand{\bc}{\boldsymbol{c}}
\newcommand{\bme}{\boldsymbol{e}}

\newcommand{\bphi}{{\boldsymbol \phi}}

\newcommand{\Tht}{{\boldsymbol \theta}}

\begin{document}
\allowdisplaybreaks

\newcommand{\arXivNumber}{2206.14403}

\renewcommand{\PaperNumber}{007}

\FirstPageHeading

\ShortArticleName{On the Fourth-Order Lattice Gel'fand--Dikii Equations}

\ArticleName{On the Fourth-Order Lattice Gel'fand--Dikii Equations}

\Author{Guesh Yfter TELA~$^{\rm a}$, Song-Lin ZHAO~$^{\rm b}$ and Da-Jun ZHANG~$^{\rm a}$}

\AuthorNameForHeading{G.Y.~Tela, S.-L.~Zhao and D.-J.~Zhang}

\Address{$^{\rm a)}$~Department of Mathematics, Shanghai University, Shanghai 200444, P.R.~China}
\EmailD{\href{mailto:gueshyftr2002@gmail.com}{gueshyftr2002@gmail.com}, \href{mailto:djzhang@staff.shu.edu.cn}{djzhang@staff.shu.edu.cn}}

\Address{$^{\rm b)}$~Department of Applied Mathematics, Zhejiang University of Technology,\\
\hphantom{$^{\rm b)}$}~Hangzhou 310023, P.R.~China}
\EmailD{\href{mailto:songlinzhao@zjut.edu.cn}{songlinzhao@zjut.edu.cn}}

\ArticleDates{Received July 02, 2022, in final form February 06, 2023; Published online February 21, 2023}

\Abstract{The fourth-order lattice Gel'fand--Dikii equations in quadrilateral form are investigated. Utilizing the direct linearization approach, we present some equations of the extended lattice Gel'fand--Dikii type. These equations are related to a quartic discrete dispersion relation and can be viewed as higher-order members of the extended lattice Boussinesq type equations. The resulting lattice equations given here are in five-component form, and some of them are multi-dimensionally consistent by introducing extra equations. Lax integrability is discussed both by direct linearization scheme and also through multi-dimensional consistent property. Some reductions of the five-component lattice equations to the four-component forms are considered.}


\Classification{35Q51; 35Q55}

\vspace{-2mm}

\section{Introduction}\label{sec-1}

The Gel'fand--Dikii (GD) hierarchy \cite{GD-1976,GD-1977} is a generalization of the Korteweg--de Vries
(KdV) hierarchy.
The GD spectral problem is known as
\begin{equation*}
L\psi=\lambda\psi,\qquad L=\partial^n_x+u_2\partial^{n-2}_x+\cdots+ u_{n-1}\partial_x + u_n,
\end{equation*}
where
$\lambda$ is a spectral parameter, $u_i=u_i(t,x)$ and $\partial_x=\frac{\partial}{\partial x}$.
When $n=2$ and $n=3$, the above spectral problem gives rise to the KdV hierarchy and Boussinesq hierarchy, respectively.
The semi-discrete (differential-difference) GD hierarchy is related to the spectral problem
\[L\psi(n)=\lambda\psi(n),\qquad L=E^m+\sum_{j=1}^{m+k} u_j(n)E^{m-j},\qquad m,k\geq 1,\quad m,k\in \mathbb{Z},\]
where $u_j(n)=u_j(n,t)$ are functions defined on $\mathbb{Z}\times \mathbb{R}$,
$E$ is the shift operator defined as $E^jf(n)=f(n+j)$ for $j\in \mathbb{Z}$.
The semi-discrete GD hierarchy has been systematically studied by Kupershmidt in~\cite{K-1985}.
Some explicit lower-order members can be found in \cite{BM-1994}.
Both continuous and semi-discrete GD hierarchies can be studied as
reductions of the Kadomtsev--Petviashvili (KP) hierarchy related to a pseudo-differential operator
or differential-difference KP hierarchy related to a pseudo-difference operator
in the frame of Sato's KP theory.

Compared with continuous and semi-discrete cases, there is no pseudo-difference operator
available for the fully discrete KP equations.
Therefore, the discrete Boussinesq (DBSQ) type equations and higher-order GD type lattice equations
are expected to play more roles
in understanding the discrete KdV/KP type integrable systems.
For the DBSQ equations, one can refer to~\cite{HZ-2021,Walker} and the references therein.
As discrete models, the lattice GD hierarchy was first introduced
by Nijhoff et al.~\cite{GD} via direct linearization (DL)
approach and transforms to the GD differential hierarchy
under continuum limits.
This hierarchy has received a lot of attention in understanding the discrete KdV/KP type equations.
In~\cite{GD-Lag}, a~Lagrangian for the generic member of the lattice GD hierarchy was presented.
This is a~Lagrangian 2-form when embedded in a higher-dimensional
lattice, obeying a closure relation. Subsequently, an integrable noncommutative
modified lattice GD hierarchy was introduced~\cite{Doli-ncmGD}. The Lax integrability and
multidimensional consistency (MDC) were shown.
Very recently, a variational perspective on continuum limits
for the lattice GD hierarchy was also discussed~\cite{Ver-CL}.\looseness=-1

The simplest member of lattice GD hierarchy is the lattice (potential) KdV equation
\begin{align}\label{eq:LPKdV}
(p-q+u_{n,m+1}-u_{n+1,m})(p+q+u_{n,m}-u_{n+1,m+1})=p^2-q^2,
\end{align}
where dependent variable $u_{n,m}$ is a function
defined on the two-dimensional lattice with discrete coordinates $(n,m)\in\mathbb{Z}^2$,
$p$ and $q$ denote lattice parameters associated with the directions in the lattice.
The equation \eqref{eq:LPKdV} first appeared as a nonlinear superposition
formula of the B\"{a}cklund transformation of the KdV equation \cite{Hugo}.
This equation, together with the lattice modified KdV equation as well as the lattice Schwarzian KdV equation,
composes the lattice KdV type equations.
Analogously, the latter two equations were, respectively, obtained as
nonlinear superposition formulas of the B\"{a}cklund transformations for the modified KdV equation and
the Schwarzian KdV equation. All these three equations can arise from the so-called Nijhoff--Quispel--Capel equation
as distinct parameter choices \cite{NQC-1983}.
Besides, all of them possess a multi-dimensionally consistent (MDC) property,
which allows a lattice equation (or a system) to be consistently embedded into a higher dimension
\cite{ABS-2003,BS-QG,Nij-LP,NW-2001}.
With this property and three additional requirements on lattice equations: affine linear,
$ D_4$ symmetry, and tetrahedron property,
Adler, Bobenko, and Suris (ABS) classified lattice models defined on an elementary quadrilateral \cite{ABS-2003}.
Some of the lattice equations in the ABS list are the lattice KdV type equations mentioned above.\looseness=-1

As a higher member in the lattice GD hierarchy, the potential DBSQ equation reads \cite{GD}
\begin{gather}
 \frac{p^3-q^3}{p-q-v_{n+1,m+1}+v_{n,m+2}}-\frac{p^3-q^3}{p-q-v_{n+2,m}+v_{n+1,m+1}} \nn \\
 \qquad{}=(p-q-v_{n+1,m}+v_{n,m+1})(2p+q-v_{n+2,m+1}+v_{n,m}) \nn \\
 \qquad\quad {}-(p-q-v_{n+2,m+1}+v_{n+1,m+2})(2p+q-v_{n+2,m+2}+v_{n,m+1}),\label{eq:LBSQ}
\end{gather}
which is defined on a $3\times 3$ stencil on the discrete two-dimensional plane. Similar to the lattice KdV case,
here dependent variable $v$ is a function of discrete independent variab\-les~$n$ and~$m$,
and~$p$ and~$q$ are lattice parameters.
Together with the potential DBSQ equation \eqref{eq:LBSQ}, also lattice versions of the modified Boussinesq
and Schwarzian Boussinesq equations have been found (see \cite{N-cof,GD,Walker}).
These equations constitute the DBSQ type equations.
By introducing two additional variables, equation \eqref{eq:LBSQ} can be expressed as a three-component form,
which is defined at the vertices of an elementary square and possesses the MDC property \cite{TN2}.
Hietarinta~\cite{H} generalised this result and gave a classification of the three-component equations
satisfying the MDC property, with some equations defined on the edges of the consistency cube and
others on the faces of the cube. It was revealed that the equations in the
classification followed one and the same
underlying structure, and some of them can be viewed as extended DBSQ type equations~\cite{ZZN}.
We refer the reader to the recent survey \cite{HZ-2021} and
the references therein for the DBSQ type equations and many interesting results.\looseness=1

The equations of each order in the extended lattice GD hierarchy
are identified by the polynomial of symmetric form
\begin{gather}\label{e-curve-N}
G_N(\omega,k):=\sum^N_{j=1}\alpha_j\big(\omega^j-k^j\big), \qquad \alpha_N\equiv 1,
\end{gather}
with coefficients $\{\alpha_j\}$ and parameter $k$. It is worthy to note that parameter $\alpha_1$ yields trivial
extension for the whole extended lattice GD type hierarchies, while
the parameters $\{\alpha_j\}_{j=2}^N$ lead to nontrivial extension (cf.~\cite{ZZN}).
Hence, $N=2$ leads to the usual
lattice KdV case and $N=3$ yields the extended DBSQ case~\cite{ZZN}.
By ``extended'' we mean the parameters $\{\alpha_j\}$
are involved in the lattice equations. In addition, more details can be found in Section~\ref{sec-2} on how \eqref{e-curve-N} is used to define
plane wave factors of the KdV case and DBSQ case.

This paper is devoted to lattice GD type equations arising from
quartic polynomial $G_4(x,k)$, by using the DL approach.
This method, based on the use of linear singular integral equations, was introduced by
Fokas and Ablowitz in order to solve the KdV equation and
treat the initial value problem of the Painlev\'e II equation \cite{FA-DL}.
Soon after the discrete DL scheme was developed~\cite{NQC-1983,QNCL}.
 And then, this method has been widely used to investigate continuous, semi-discrete,
 and discrete integrable systems \cite{Fu-2,Fu-3,Fu-1,Fu-4,1995-dKdV,NQC-1983,QNCL,SAF,Fu-5,ZZN}.
In this paper, we are interested in the construction of the fourth-order lattice GD type equations as well as their
MDC property and Lax integrability. For the sake of brevity, we call the resulting
lattice equations as lattice GD-4 type equations.
We will see that the lattice GD-4 type equations exhibit many nontrivial features compared with the DBSQ type equations.

The paper is arranged as follows: In Section~\ref{sec-2}, we give a brief review of the
DL scheme for the extended lattice GD type equations. In Section~\ref{sec-3}, we
construct the lattice GD-4 type equations as closed-forms, which are
transformed into deformed forms in Section~\ref{sec-4}. Then, Section~\ref{sec-5} is devoted to the
MDC property and Lax representation, and Section~\ref{sec-6} is for conclusions.
There are three appendices, where we list the DBSQ equations, present details of
some calculations and triply shifted quantities of the variables from MDC
of some lattice GD-4 equations, respectively.

\section{Framework of discrete direct linearisation approach}\label{sec-2}

In this section, we briefly recall the DL framework for the extended lattice GD type equations
outlined in the paper \cite{ZZN}.
Let us introduce the following conventional short hands
\[f=f(n,m), \qquad \wt{f}=f(n+1,m), \qquad \wh{f}=f(n,m+1), \qquad \wh{\wt{f}}=f(n+1,m+1).\]
Besides, we employ $\oa_j(k)$, $j=1,2,\dots,N-1$, and $\oa_{N}(k)=k$
to represent the roots of equation $G_N(\omega,k)=0$ given in~\eqref{e-curve-N}.

The starting point in the DL approach is the following $\infty\times\infty$ matrix $\bC$ defined by the integral
\begin{align}
\label{eq:bC}
\bC=\sum_{j=1}^N \oint_{\Gamma_j}\,{\rm d}\ld_j(k)\,\rho_k
\bc_k \tc_{-\oa_j(k)}\sg_{-\oa_j(k)},\qquad N\geq 2,
\end{align}
where $\{{\rm d}\ld_j(k)\}$ are certain measures for the integral on integration contours $\Gamma_j$ in the
space of the spectral variable $k$, $\bc_k=(k^j)_{j\in\mathbb{Z}}$ stands for the column vector\footnote{We use ``${\rm t}$'' to denote the transpose of a matrix with infinite dimensions, e.g.,
$\tc$, $\tLd$, and use ``${\rm T}$'' for the finite-dimensional case, e.g., equation~\eqref{5.3}.}
$\big(\dots, k^{-2}, k^{-1}, 1, k, k^2, \dots\big)^{\rm T}$,
and $\tc_{k'}$ denotes the transpose of~$\bc_{k'}$.
The factors $\rho_k$, $\sg_{k'}$ are discrete exponential functions of~$k$ and $k'$, respectively, given by
\begin{align*}
\rho_k=(p+k)^n(q+k)^m\rho^0_k,\qquad \sg_{k'}=\big(p-k'\big)^{-n} (q-k' )^{-m}\sg^0_{k'},
\end{align*}
where $\rho^0_{k}, \sg^0_{k'}\in \mathbb{C}$.
We assume that basic operations (such as differentiations with respect to parameters or applying shifts in the
variables $n$ and $m$) commute with the integrations.
Note that $\rho_{k}$ and $\sg_{-\oa_j(k)}$ can be considered to be defined
using the roots of $G_N(-p,k)=0$. When $N=2$, we have
\[
\rho=\rho_{k} \sg_{-\oa_j(k)}=\left(\frac{p+k}{p-k}\Bigr)^n \Bigl(\frac{q+k}{q-k}\right)^m,
\]
which leads to the lattice KdV type equations, cf.~\cite{QNCL}.
When $N=3$, we have
\[
\rho=\rho_{k} \sg_{-\oa_j(k)}=\left(\frac{p+k}{p+\omega_j(k)}\right)^n
\left(\frac{q+k}{q+\omega_j(k)}\right)^m,
\]
 for $j=1,2$,
which leads to the extended DBSQ type equations, see more details in \cite{ZZN}
for how these equations are derived in DL approach.
Note also that by ``extended'' we mean in the lattice equations the parameters $\{\alpha_j\}$ are involved.

By introducing $\infty\times\infty$ matrices $\Ld$ and $\tLd$ defined by
\begin{align*}
\Ld\,\bc_k=k\,\bc_k,\qquad \tc_{k'}\tLd=k'\,\tc_{k'},
\end{align*}
one easily identify that matrix $\bC$ obeys the linear relations \cite{ZZN}
\begin{subequations}
\label{eq:bCdyn}
\begin{gather}
\wt{\bC} \big(p-\tLd\big)=(p+\Ld) \bC,\qquad \wh{\bC} \big(q-\tLd\big)=(q+\Ld) \bC, \\
\label{eq:idens_b}
\left[\prod_{j=1}^{N}\big(\oa_j(-p)-\Ld\big)\right] \bC=\bC \prod_{j=1}^{N}\big(\oa_j(-p)+\tLd\big), \\
\label{eq:idens_c}
\left[\prod_{j=1}^{N-1}\big(\oa_j(-p)-\Ld\big)\right] \wt{\bC}=\bC \prod_{j=1}^{N-1}\big(\oa_j(-p)+\tLd\big),
\end{gather}
\end{subequations}
and similar relations with $p$ replaced by $q$ and $\wt{\bC}$ by $\wh{\bC}$.
To proceed, we introduce some auxiliary elements:
\begin{itemize}\itemsep=0pt
\item
a Cauchy kernel $\bOm$ defined by the relation $\bOm\Ld+\tLd\bOm=\bO$ with $\bO$ being a rank 1 projection matrix, obeying $\bO^2=\bO$,
\item
 an $\infty\times\infty$ matrix~$\bU$~obeying the relation~$\bU=\bC-\bU \bOm \bC$,
\item
 an $\infty\times 1$ vector~$\bu_k$~defined by~$\bu_k=\rho_k(\bc_k-\bU \bOm \bc_k)$.
\end{itemize}
With these elements, from \eqref{eq:bCdyn} one has
\begin{subequations}
\label{eq:NUrels}
\begin{align}
\label{eq:NUrels_a}
&\wt{\bU} \big(p-\tLd\big)=(p+\Ld)\bU-\wt{\bU} \bO \bU, \\
\label{eq:NUrels_b}
& \bU\,\left[ \prod_{j=1}^{N-1}\big(\oa_j(-p)+\tLd\big)\right]=\left[\prod_{j=1}^{N-1}\big(\oa_j(-p)-\Ld\big)\right]\wt{\bU} \nn \\
& \qquad {}+\bU\, \sum_{j=0}^{N-2} \left[ \prod_{l=1}^j
\big(\oa_l(-p)+\tLd\big)\right]\bO \left[\prod_{l=j+2}^{N-1}
\big(\oa_l(-p)-\Ld\big)\right] \wt{\bU}, \\
& \bU \sum_{j=1}^{N}\alpha_j\big({-}\tLd\big)^{j}=\sum_{j=1}^{N}\alpha_j\Ld^{j} \bU
-\bU \sum_{j=1}^{N} \alpha_j\sum_{l=0}^{j-1}
\big({-}\tLd\big)^l\bO\Ld^{j-1-l} \bU, \label{eq:NUrels_c}
\end{align}
\end{subequations}
as well as
\begin{subequations}
\label{eq:Nukrels}
\begin{align}
\label{eq:Nukrels_a}
& \wt{\bu}_k=(p+\Ld)\bu_k-\wt{\bU}\,\bO\,\bu_k, \\
\label{eq:Nukrels_b}
& -\left[ \prod_{j=1}^{N}\big(\oa_j(-p)-k\big)\right]\bu_k=\left[\prod_{j=1}^{N-1}\big(\oa_j(-p)-\Ld\big)\right]\wt{\bu}_k \nn \\
& \qquad {}+\bU\, \sum_{j=0}^{N-2} \left[ \prod_{l=1}^j
\big(\oa_l(-p)+\tLd\big)\right]\bO \left[\prod_{l=j+2}^{N-1}
\big(\oa_l(-p)-\Ld\big)\right]\,\wt{\bu}_k, \\
\label{eq:Nukrels_c}
& \sum_{j=1}^{N} \alpha_j k^{j}\bu_k=\sum_{j=1}^{N}
\alpha_j\Ld^{j}\,\bu_k -\bU\,\sum_{j=1}^{N} \alpha_j\sum_{l=0}^{j-1}
\big({-}\tLd\big)^l\bO\Ld^{j-1-l}\,\bu_k.
\end{align}
\end{subequations}
Similar relations to \eqref{eq:NUrels_a}, \eqref{eq:NUrels_b}, and
\eqref{eq:Nukrels_a} and \eqref{eq:Nukrels_b} can be obtained with
$(\wh{\phantom{a}}, q)$ to replace~$(\wt{\phantom{a}},p)$.

Relations~\eqref{eq:NUrels_a} and~\eqref{eq:NUrels_b}, together with their $(\wh{\phantom{a}},q)$ analogues,
supply important
information for the construction of lattice GD type equations.
Note that although relation~\eqref{eq:NUrels_c}
does not exert any role in the construction of discrete equations (e.g.,~\cite{FZZ-JNMP,ZZ-SAPM,ZZN})
and semi-discrete equations (e.g.,~\cite{Mes}),
it is indispensable in the construction of continuous equations (e.g.,~\cite{XZZ}).
Relations~\eqref{eq:Nukrels} are linear with respect to $\bu_k$ and usually used,
 together with their $(\wh{\phantom{a}},q)$ analogues, to construct Lax pairs
for the derived lattice equations (e.g.,~\cite{ZZN}).

\section{Lattice GD-4 type equations as quadrilateral closed forms}\label{sec-3}

In this section, we will restrict our attention to
the relations \eqref{eq:NUrels} and \eqref{eq:Nukrels} in the case of $N=4$, and
construct the lattice GD-4 type equations as quadrilateral closed forms.
In this case, $\omega_j(k)$, $j=1,2,3,4$, in \eqref{eq:NUrels} and \eqref{eq:Nukrels}
are the roots of $G_4(\omega,k)=0$, where $G_4(\omega,k)$ is the polynomial defined as in~\eqref{e-curve-N}.
In the following, we show step by step
how quadrilateral closed forms of scalar functions arise from~\eqref{eq:NUrels} and~\eqref{eq:Nukrels}.

\subsection{Basic objects and their relations}\label{sec-3-1}

When $N=4$ the set of relations \eqref{eq:NUrels} and \eqref{eq:Nukrels} take the form
\begin{subequations}
\label{eq:Urels}
\begin{align}
\label{eq:Urels_a}
& \wt{\bU} \big(p-\tLd\big) = (p+\Ld)\bU-\wt{\bU} \bO \bU, \\
\label{eq:Urels_b}
& \bU \big(A_3(-p)+A_2(-p)\tLd+A_1(-p)\tLd^2+\tLd^3\big) \nn \\
& \qquad {}=\big(A_3(-p)-A_2(-p)\Ld+A_1(-p)\Ld^2-\Ld^3\big)\wt{\bU}+\bU\big[
A_2(-p)\bO \nn \\
& \qquad\quad{} -A_1(-p)\big(\bO\,\Ld-\tLd\,\bO\big)+\big(\bO\,\Ld^2-\tLd \bO \Ld+\tLd^2 \bO\big)\big]\wt{\bU}, \\
\label{eq:Urels_c}
& \bU \sum_{j=1}^{4}\alpha_j\big({-}\tLd)^{j}=
\sum_{j=1}^{4}\alpha_j\Ld^{j} \bU -\bU \sum_{j=1}^{4}
\alpha_j\sum_{l=0}^{j-1} \big({-}\tLd)^l\bO\Ld^{j-1-l} \bU,
\end{align}
\end{subequations}
and
\begin{subequations}
\label{eq:ukrels}
\begin{align}
\label{eq:ukrels_a}
& \wt{\bu}_k = (p+\Ld)\bu_k-\wt{\bU} \bO \bu_k, \\
\label{eq:ukrels_b}
& -\left[\prod_{j=1}^{4}\big(\oa_j(-p)-k\big)\right]\bu_k =\big(A_3(-p)-A_2(-p)\Ld+A_1(-p)\Ld^2-\Ld^3\big)\wt{\bu}_k \nn \\
& \qquad{}+\bU \big[A_2(-p)\bO-A_1(-p)\big(\bO \Ld-\tLd \bO\big)
+\big(\bO \Ld^2-\tLd \bO \Ld+\tLd^2 \bO\big)\big]\wt{\bu}_k, \\
\label{eq:ukrels_c}
& \sum_{j=1}^{4} \alpha_j k^{j}\bu_k = \sum_{j=1}^{4}
\alpha_j\Ld^{j} \bu_k -\bU \sum_{j=1}^{4} \alpha_j\sum_{l=0}^{j-1}
\big({-}\tLd\big)^l\bO\Ld^{j-1-l} \bu_k,
\end{align}
\end{subequations}
and these relations hold when ($\wt{\phantom{a}}, p)$ is replaced by $(\wh{\phantom{a}}, q)$.
Here $\{A_j(k)\}$ are defined as
\begin{align*}
& A_1(k)=\oa_1(k)+\oa_2(k)+\oa_3(k)=-(k+\alpha_3), \\
& A_2(k)=\oa_1(k)\oa_2(k)+\oa_1(k)\oa_3(k)+\oa_2(k)\oa_3(k)=k^2+\alpha_3 k+\alpha_2, \\
& A_3(k)=\oa_1(k)\oa_2(k)\oa_3(k)=-\big(k^3+\alpha_3 k^2+\alpha_{2}k+\alpha_{1}\big),
\end{align*}
and $\{\oa_j(k)\}$ are the roots of $G_4(\omega,k)=0$, i.e.,
\[G_4(\omega,k)=\prod_{j=1}^{4}(\omega-\oa_j(k)),\qquad \oa_4(k)=k.\]

Next, in order to construct the lattice equations, let us introduce the following scalar objects:%
\begin{subequations}\label{eq:objs}
\begin{align}
& v_a: = 1-\tbme\,(a+\Ld)^{-1} \bU \bme,\qquad w_b:= 1+\tbme \bU \big(b - \! \tLd\big)^{-1} \bme, \\
& s_a: = a-\tbme\,(a+\Ld)^{-1} \bU \tLd \bme,\quad t_b:= -b+\tbme \Ld \bU \big(b - \! \tLd\big)^{-1} \bme, \\
& r_a: = a^2-\tbme\,(a+\Ld)^{-1}\,\bU\,\tLd^2\,\bme,\quad
z_b:= b^2+\tbme \Ld^2 \bU \big(b - \! \tLd\big)^{-1} \bme, \\
& f_a: = a^3-\tbme (a+\Ld)^{-1} \bU \tLd^3 \bme,\qquad
g_b:= -b^3+\tbme \Ld^3 \bU \big(b - \! \tLd\big)^{-1} \bme, \\
& u_{i,j}:= \tbme\Ld^i\bU\tLd^j\bme, \qquad
s_{a,b}:= \tbme (a+\Ld)^{-1} \bU \big({-}b+ \! \tLd\big)^{-1}\bme,\qquad i,j\in\mathbb{Z}.
\end{align}
\end{subequations}
Here $a$ and $b$ are parameters,
$\bme$ is a fixed column vector $(\dots,0,0,1,0,0,\dots)^{\rm T}$ where the only nonzero entry $1$ is in the central,
the adjoint vector $\tbme$ is the transpose of $\bme$. It then turns out that\looseness=-1
\begin{gather*}
\tbme \bc_k=1, \qquad \tc_{k'} \bme=1, \qquad \bO=\bme \tbme.
\end{gather*}
Except $f_a$ and $g_b$, all others take the same forms as in \cite{ZZN} for the DBSQ case.
Based on the shift relations~\eqref{eq:Urels_a} and~\eqref{eq:Urels_b}, for the objects~\eqref{eq:objs} we have the following relations:
multiplying~\eqref{eq:Urels_a} and~\eqref{eq:Urels_b}
from the left by $\tbme$ and from the right by $\bme$, we have
\begin{subequations}
\label{eq:uij}
\begin{gather}
\label{eq:uij_a}
 p\wt{u}_{i,j}-\wt{u}_{i,j+1}=pu_{i,j}+u_{i+1,j}-\wt{u}_{i,0} u_{0,j}, \\
A_3(-p)u_{i,j}+A_2(-p)u_{i,j+1}+A_1(-p)u_{i,j+2}+u_{i,j+3}\nn \\
\qquad {}= A_3(-p)\wt{u}_{i,j}-A_2(-p)\wt{u}_{i+1,j}+A_1(-p)\wt{u}_{i+2,j}-\wt{u}_{i+3,j} \nn \\
 \qquad\quad{} +A_2(-p)u_{i,0}\wt{u}_{0,j}+A_1(-p)\big(u_{i,1}\wt{u}_{0,j}-u_{i,0}\wt{u}_{1,j}\big)+
u_{i,0}\wt{u}_{2,j}-u_{i,1}\wt{u}_{1,j}+u_{i,2}\wt{u}_{0,j},\!\!\!\label{eq:uij_b}
\end{gather}
\end{subequations}
where $p_a$, $p_b$ are defined as
\begin{subequations}
\label{eq:papb}
\begin{gather}
 p_a=G_4(-p,-a)/(p-a)=A_3(-p)+aA_2(-p)+a^2A_1(-p)+a^3, \\
 p_b=G_4(-p,-b)/(p-b)=A_3(-p)+bA_2(-p)+b^2A_1(-p)+b^3;
\end{gather}
\end{subequations}
we can also derive
(as examples of derivation, we provide details in Appendix \ref{App-2} for
equations \eqref{eq:ssrels_b}, \eqref{eq:fgrels_a} and \eqref{eq:gfrels_a})
\begin{subequations}
\label{eq:ssrels}
\begin{gather}
\label{eq:ssrels_a}
 1+(p-a)s_{a,b}-(p-b)\wt{s}_{a,b}=\wt{v}_a w_b, \\
 \big[A_2(-p)+(a+b)A_1(-p)+a^2+ab+b^2\big]+p_bs_{a,b}-p_a\wt{s}_{a,b} \nn \\
\qquad{} = v_a\wt{z}_b-s_a \wt{t}_b+r_a\wt{w}_b+A_2(-p)v_a
\wt{w}_b-A_1(-p)\big(v_a \wt{t}_b-s_a \wt{w}_b\big),\label{eq:ssrels_b}
\end{gather}
\end{subequations}
and
\begin{subequations}
\label{eq:strels}
\begin{gather}
\label{eq:strels_a}
\wt{s}_a=(p+u_0)\wt{v}_a-(p-a) v_a, \\
\label{eq:strels_b}
 t_b=(p-b)\wt{w}_b-(p-\wt{u}_0)w_b,
\end{gather}
\end{subequations}
and
\begin{subequations}
\label{eq:zrrels}
\begin{gather}
\label{eq:zrrels_a}
 \wt{r}_a=p\wt{s}_a-(p-a) s_a+\wt{v}_a u_{0,1}, \\
\label{eq:zrrels_b}
 z_b=(p-b) \wt{t}_b-p t_b+\wt{u}_{1,0} w_b,
\end{gather}
\end{subequations}
and
\begin{subequations}
\label{eq:fgrels}
\begin{gather}
\label{eq:fgrels_a}
 \wt{f}_a=p\wt{r}_a-(p-a) r_a+\wt{v}_a u_{0,2}, \\
\label{eq:fgrels_b}
 g_b=(p-b) \wt{z}_b-p z_b+\wt{u}_{2,0} w_b,
\end{gather}
\end{subequations}
and
\begin{subequations}
\label{eq:gfrels}
\begin{gather}
 f_a = p_a\wt{v}_a-\big[A_3(-p)-A_2(-p)\wt{u}_{0}+A_1(-p)\wt{u}_{1,0}-\wt{u}_{2,0}\big]v_a \nn \\
\hphantom{f_a =}{}
 -\big[A_2(-p)-A_1(-p)\wt{u}_{0}+\wt{u}_{1,0}\big]s_a-\big(A_1(-p)-\wt{u}_0\big)r_a, \label{eq:gfrels_a}\\
 \wt{g}_b =-p_b w_b+[A_3(-p)+A_2(-p)u_{0}+A_1(-p)u_{0,1}+u_{0,2}]\wt{w}_b \nn \\
\hphantom{\wt{g}_b =}{} -[A_2(-p)+A_1(-p)u_{0}+u_{0,1}]\wt{t}_b+(A_1(-p)+u_0)\wt{z}_b,\label{eq:gfrels_b}
\end{gather}
\end{subequations}
where we have used a simplified
notation $u_0=u_{0,0}$.
All these relations, \eqref{eq:ssrels}--\eqref{eq:gfrels}, also
hold for their hat-$q$ counterparts obtained by replacing $(\wt{\phantom{a}}, p)$ with $(\wh{\phantom{a}},q)$.
It should be emphasized that
$q_a$ and $q_b$ appear in the rest part of the paper have the same form with $p_a$ and $p_b$
just with replacement $p\rightarrow q$.

A further set of relations is
\begin{subequations}
\label{eq:svurels}
\begin{gather}
\big[H(p,q)+\wh{\wt{u}}_{1,0}-(p+q-\alpha_3)\wh{\wt{u}}_{0}
+ \big((p+q-\alpha_3-\wh{\wt{u}}_{0})s_a+r_a\big)/v_a\big] \big(p-q+\wh{u}_0-\wt{u}_0\big)\nn \\
 \qquad =\big(p_a\wt{v}_a-q_a\wh{v}_a\big)/v_a,\label{eq:svurels_a} \\
 p-q+\wh{u}_0-\wt{u}_0= \big[(p-a)\wh{v}_a-(q-a)\wt{v}_a\big]/\wh{\wt{v}}_a, \label{eq:svurels_b}\\
 \big(\wt{r}_a-p\wt{s}_a\big)/\wt{v}_a-\big(\wh{r}_a-q\wh{s}_a\big)/\wh{v}_a=-(p-a)s_a/\wt{v}_a+(q-a)s_a/\wh{v}_a,
\label{eq:svurels_c}
\end{gather}
\end{subequations}
which are derived from equations \eqref{eq:uij_a}, \eqref{eq:strels_a}, \eqref{eq:zrrels_a},
\eqref{eq:gfrels_a} and their hat-$q$ counterparts,
where
\[H(p,q)=(A_3(-p)-A_3(-q))/(p-q).\]
Similarly, from \eqref{eq:uij_a}, \eqref{eq:strels_b}, \eqref{eq:zrrels_b}, \eqref{eq:gfrels_b}
and their hat-$q$ versions one has
\begin{subequations}
\label{eq:twurels}
\begin{gather}
\label{eq:twurels_a}
 \big[H(p,q)+u_{0,1}+(p+q-\alpha_3)u_{0}
- \big((p+q-\alpha_3+u_{0})\wh{\wt{t}}_b-\wh{\wt{z}}_b\big)/\wh{\wt{w}}_b\big] \nn \\
 \qquad{}\times \big(p-q+\wh{u}_0-\wt{u}_0\big)=\big(p_b\wh{w}_b-q_b\wt{w}_b\big)/\wh{\wt{w}}_b, \\
\label{eq:twurels_b}
 p-q+\wh{u}_0-\wt{u}_0= [(p-b)\wt{w}_b-(q-b)\wh{w}_b]/w_b, \\
 \big(\wh{z}_b+p\wh{t}_b\big)/\wh{w}_b-\big(\wt{z}_b+q\wt{t}_b\big)/\wt{w}_b
=(p-b)\wh{\wt{t}}_b/\wh{w}_b-(q-b)\wh{\wt{t}}_b/\wt{w}_b.
\label{eq:twurels_c}
\end{gather}
\end{subequations}

\subsection{Lattice equations as quadrilateral closed forms}\label{sec-3-2}

We have achieved many relations with respect to the scalar elements defined in equation \eqref{eq:objs}.
With these relations in hand, we are able to derive quadrilateral closed forms of some elements,
 which can be considered as lattice equations.

We start by presenting a set of lattice equations, which are made up of variables $v_a$,~$s_a$,~$r_a$,~$u_0$ and $u_{1,0}$.
To do that, we first set $i=j=0$ in equation \eqref{eq:uij_a} and its hat-$q$ counterpart,
which yield
\begin{align}
\label{eq:u00-eq1}
p\wt{u}_{0}-\wt{u}_{0,1}=pu_{0}+u_{1,0}-\wt{u}_{0} u_{0},\qquad
q\wh{u}_{0}-\wh{u}_{0,1}=qu_{0}+u_{1,0}-\wh{u}_{0} u_{0}.
\end{align}
It then follows that
\begin{subequations}
\label{eq:u00-10-01}
\begin{align}
\label{eq:u00-10}
& \wh{u}_{1,0}-\wt{u}_{1,0}=\big(p-q+\wh{u}_0-\wt{u}_0\big)\wh{\wt{u}}_0-p\wh{u}_0+q\wt{u}_0, \\
\label{eq:u00-01}
& \wh{u}_{0,1}-\wt{u}_{0,1}=\big(p-q+\wh{u}_0-\wt{u}_0\big)u_0-p\wt{u}_0+q\wh{u}_0.
\end{align}
\end{subequations}
Obviously, equation \eqref{eq:strels_a} and its hat-$q$ counterpart, \eqref{eq:svurels_a},
\eqref{eq:svurels_c} and \eqref{eq:u00-10} give rise to a~closed-form with respect to $v_a$, $s_a$, $r_a$, $u_0$ and $u_{1,0}$,
i.e.,
\begin{subequations}
\label{eq:A2-1}
\begin{gather}
\label{eq:bA2-a}
 \wt{s}_a=(p+u_0)\wt{v}_a-(p-a) v_a, \quad \wh{s}_a=(q+u_0)\wh{v}_a-(q-a) v_a,\\
\label{eq:bA2-b}
 \big(\wt{r}_a-p\wt{s}_a\big)/\wt{v}_a-\big(\wh{r}_a-q\wh{s}_a\big)/\wh{v}_a=-(p-a)s_a/\wt{v}_a+(q-a)s_a/\wh{v}_a, \\
\label{eq:bA2-c}
 \wh{u}_{1,0}-\wt{u}_{1,0}=\big(p-q+\wh{u}_0-\wt{u}_0\big)\wh{\wt{u}}_0-p\wh{u}_0+q\wt{u}_0, \\
 \big[H(p,q)+\wh{\wt{u}}_{1,0}-(p+q-\alpha_3)\wh{\wt{u}}_{0}
+\big(\big(p+q-\alpha_3-\wh{\wt{u}}_{0}\big)s_a+r_a\big)/v_a\big] \nn \\
 \qquad {}\times \big(p-q+\wh{u}_0-\wt{u}_0\big)=\big(p_a\wt{v}_a-q_a\wh{v}_a\big)/v_a.\label{eq:bA2-d}
\end{gather}
\end{subequations}
This is a set of equations consisting of five equations of five scalar functions.
We will revisit it in Section~\ref{sec-4}.
A similar closed-form is composed by \eqref{eq:strels_b}, its hat-$q$ counterpart,
\eqref{eq:twurels_a}, \eqref{eq:twurels_c} and \eqref{eq:u00-01}, i.e.,
\begin{subequations}
\label{eq:A2-2}
\begin{gather}
\label{eq:A2-b}
 t_b =(p-b)\wt{w}_b-(p-\wt{u}_0)w_b, \quad t_b= (q-b)\wh{w}_b-(q-\wh{u}_0)w_b, \\
\label{eq:bA2-c-b}
 (\wh{z}_b+p\wh{t}_b)/\wh{w}_b-(\wt{z}_b+q\wt{t}_b)/\wt{w}_b
=(p-b)\wh{\wt{t}}_b/\wh{w}_b-(q-b)\wh{\wt{t}}_b/\wt{w}_b, \\
\label{eq:bA2-c-c}
 \wh{u}_{0,1}-\wt{u}_{0,1}=(p-q+\wh{u}_0-\wt{u}_0)u_0-p\wt{u}_0+q\wh{u}_0, \\
 \big[H(p,q)+u_{0,1}+(p+q-\alpha_3)u_{0}
- \big((p+q-\alpha_3+u_{0})\wh{\wt{t}}_b-\wh{\wt{z}}_b\big)/\wh{\wt{w}}_b\big] \nn \\
 \qquad {}\times (p-q+\wh{u}_0-\wt{u}_0)=(p_b\wh{w}_b-q_b\wt{w}_b)/\wh{\wt{w}}_b,\label{eq:bA2-c-d}
\end{gather}
\end{subequations}
which are equations of $t_b$, $w_b$, $z_b$, $u_{0,1}$ and $u_{0}$.

There are more closed forms.
Taking $i=j=0$ in \eqref{eq:uij_b} and its hat-$q$ counterpart yields
\begin{subequations}
\label{eq:u00-eq2}
\begin{gather}
 A_3(-p)u_0+A_2(-p)u_{0,1}+A_1(-p)u_{0,2}+u_{0,3}\nn\\
 \qquad {}=A_3(-p)\wt{u}_0-A_2(-p)\wt{u}_{1,0}+A_1(-p)\wt{u}_{2,0}-\wt{u}_{3,0} \nn \\
 \qquad \quad{} +A_2(-p)u_0\wt{u}_0+A_1(-p)\big(u_{0,1}\wt{u}_0-u_0\wt{u}_{1,0}\big)
+u_0\wt{u}_{2,0}-u_{0,1}\wt{u}_{1,0}+u_{0,2}\wt{u}_0,\label{eq:u00-eq2_a} \\
 A_3(-q)u_0+A_2(-q)u_{0,1}+A_1(-q)u_{0,2}+u_{0,3}\nn\\
 \qquad =A_3(-q)\wh{u}_0-A_2(-q)\wh{u}_{1,0}+A_1(-q)\wh{u}_{2,0}-\wh{u}_{3,0} \nn \\
 \qquad \quad{} +A_2(-q)u_0\wh{u}_0+A_1(-q)\big(u_{0,1}\wh{u}_0-u_0\wh{u}_{1,0}\big)
+u_0\wh{u}_{2,0}-u_{0,1}\wh{u}_{1,0}+u_{0,2}\wh{u}_0.\label{eq:u00-eq2_b}
\end{gather}
\end{subequations}
We further set $(i,j)=(1,0)$ and $(0,1)$ in \eqref{eq:uij_a} and obtain
\begin{subequations}\label{eq:u10-01-1}
\begin{gather}
\label{eq:u10-1}
 p\wt{u}_{1,0}-\wt{u}_{1,1}=pu_{1,0}+u_{2,0}-\wt{u}_{1,0} u_0,
\qquad q\wh{u}_{1,0}-\wh{u}_{1,1}=qu_{1,0}+u_{2,0}-\wh{u}_{1,0} u_0, \\
\label{eq:u01-1}
 p\wt{u}_{0,1}-\wt{u}_{0,2}=pu_{0,1}+u_{1,1}-\wt{u}_0 u_{0,1},
\qquad q\wh{u}_{0,1}-\wh{u}_{0,2}=qu_{0,1}+u_{1,1}-\wh{u}_0 u_{0,1}.
\end{gather}
\end{subequations}
Eliminating $u_{1,1}$ in \eqref{eq:u10-1} and \eqref{eq:u01-1}, respectively, we reach to
\begin{subequations}
\label{eq:u11-can}
\begin{gather}
\label{eq:u11-can_a}
 \wh{u}_{2,0}-\wt{u}_{2,0}=-p\wh{u}_{1,0}+q\wt{u}_{1,0}+\big(p-q+\wh{u}_0-\wt{u}_0\big)\wh{\wt{u}}_{1,0}, \\
\label{eq:u11-can_b}
 \wh{u}_{0,2}-\wt{u}_{0,2}=-p\wt{u}_{0,1}+q\wh{u}_{0,1}+\big(p-q+\wh{u}_0-\wt{u}_0\big)u_{0,1}.
\end{gather}
\end{subequations}
Next, subtracting \eqref{eq:u00-eq2_b} from \eqref{eq:u00-eq2_a} yields an equation without
$u_{0,3}$ but containing $u_{3,0}$.
To delete $u_{3,0}$, we make use of
equations \eqref{eq:u11-can_a} and \eqref{eq:uij_a} with $(i,j)=(2,0)$ and $(1,0)$.
Then, after some algebra we arrive at{\samepage
\begin{gather}
\frac{G_4(-p,-q)}{p-q+\wh{u}_0-\wt{u}_0}=\frac{G_4(-p,-q)}{p-q}+H(p,q)\big(u_0-\wh{\wt{u}}_0\big) \nn \\
\qquad{}-(p+q-\alpha_3)\big(u_0\wh{\wt{u}}_0-\wh{\wt{u}}_{1,0}+u_{0,1}\big)+u_0\wh{\wt{u}}_{1,0}
-\wh{\wt{u}}_0u_{0,1}-\wh{\wt{u}}_{2,0}+u_{0,2}.\label{eq:BSQ-eq3}
\end{gather}}

\pagebreak

\noindent
Equation \eqref{eq:u00-eq1} and \eqref{eq:u11-can} together with \eqref{eq:BSQ-eq3}
compose a new closed-form with respect to $u_0$, $u_{0,1}$, $u_{1,0}$, $u_{0,2}$ and $u_{2,0}$, i.e.,
\begin{subequations}
\label{eq:B2}
\begin{gather}
 p\wt{u}_{0}-\wt{u}_{0,1}=pu_{0}+u_{1,0}-\wt{u}_{0} u_{0},\qquad
q\wh{u}_{0}-\wh{u}_{0,1}=qu_{0}+u_{1,0}-\wh{u}_{0} u_{0},\\
 \wh{u}_{2,0}-\wt{u}_{2,0}=-p\wh{u}_{1,0}+q\wt{u}_{1,0}+\big(p-q+\wh{u}_0-\wt{u}_0\big)\wh{\wt{u}}_{1,0}, \\
 \wh{u}_{0,2}-\wt{u}_{0,2}=-p\wt{u}_{0,1}+q\wh{u}_{0,1}+\big(p-q+\wh{u}_0-\wt{u}_0\big)u_{0,1},\\
 \frac{G_4(-p,-q)}{p-q+\wh{u}_0-\wt{u}_0}=\frac{G_4(-p,-q)}{p-q}+H(p,q)\big(u_0-\wh{\wt{u}}_0\big) \nn \\
 \qquad{}
-(p+q-\alpha_3)\big(u_0\wh{\wt{u}}_0-\wh{\wt{u}}_{1,0}+u_{0,1}\big)+u_0\wh{\wt{u}}_{1,0}
-\wh{\wt{u}}_0u_{0,1}-\wh{\wt{u}}_{2,0}+u_{0,2}.
\end{gather}
\end{subequations}

Finally, we look for a closed-form containing the element $s_{a,b}$.
By transformation
\begin{gather*}
s_{a,b}=S_{a,b}-1/(b-a), \qquad a\neq b,
\end{gather*}
the relation \eqref{eq:ssrels} and its hat-$q$ counterpart are rewritten as
\begin{subequations} \label{eq:SSrels}
\begin{gather}
\label{eq:SSrels_ab}
 (p-a)S_{a,b}-(p-b)\wt{S}_{a,b}=\wt{v}_aw_b, \qquad (q-a)S_{a,b}-(q-b)\wh{S}_{a,b}=\wh{v}_a w_b, \\
\label{eq:SSrels_c}
 p_bS_{a,b}-p_a\wt{S}_{a,b}=A_2(-p)v_a\wt{w}_b-A_1(-p)\big(v_a \wt{t}_b
-s_a \wt{w}_b\big)+v_a \wt{z}_b-s_a \wt{t}_b+r_a \wt{w}_b, \\
\label{eq:SSrels_d}
 q_bS_{a,b}-q_a\wh{S}_{a,b}=A_2(-q)v_a\wh{w}_b-A_1(-q)\big(v_a \wh{t}_b
-s_a \wh{w}_b\big)+v_a \wh{z}_b-s_a \wh{t}_b+r_a \wh{w}_b.
\end{gather}
\end{subequations}
Deleting $\wh{\wt{z}}_{b}$ in $\wh{\eqref{eq:SSrels_c}}$ (the hat-shifted \eqref{eq:SSrels_c})
by utilizing \eqref{eq:twurels_a} and deleting $\wh{s}_a$ by
using the hat-$q$ version of \eqref{eq:strels_a} and also making use
of \eqref{eq:twurels_b} we get
\begin{gather}
(q-a)\big(v_{a}\wh{\wt{t}}_{b}-s_a\wh{\wt{w}}_b\big)=\frac{-\wh{v}_{a}w_b(p_b\wh{w}_b-q_b\wt{w}_b)}
{(p-b)\wt{w}_b-(q-b)\wh{w}_b} \nonumber\\
\qquad{}+(q-a)(p+q-\alpha_3)v_a\wh{\wt{w}}_b+p_b\wh{S}_{a,b}-p_a\wh{\wt{S}}_{a,b}.\label{eq:vwSrels}
\end{gather}
Making use of \eqref{eq:SSrels_ab} again, equation \eqref{eq:vwSrels} goes to the following desired form
\begin{gather}
\label{eq:vwSrels-1}
v_{a}\wh{\wt{t}}_{b}-s_a\wh{\wt{w}}_{b}=-w_b\frac{\frac{p_b}{p-a}\wh{w}_b\wt{v}_a
-\frac{q_b}{q-a}\wt{w}_b\wh{v}_a}{(p-b)\wt{w}_b-(q-b)\wh{w}_b}
+(p+q-\alpha_3)v_a\wh{\wt{w}}_{b}+\frac{G_4(-a,-b)}{(p-a)(q-a)}\wh{\wt{S}}_{a,b}.
\end{gather}
This procedure is not apparent. Let add some details.
First, we rewrite \eqref{eq:vwSrels} as
\begin{gather}
v_{a}\wh{\wt{t}}_{b}-s_a\wh{\wt{w}}_b= \frac{\frac{q_b\wh{v}_a \wt{w}_b}{q-a}
-\frac{p_b\wt{v}_a \wh{w}_b}{p-a}}
{(p-b)\wt{w}_b-(q-b)\wh{w}_b} w_b+\frac{\frac{p_b\wt{v}_a \wh{w}_b}{p-a}-\frac{p_b\wh{v}_a\wh{w}_b}{q-a}}
{(p-b)\wt{w}_b-(q-b)\wh{w}_b}w_b \nn\\
\hphantom{v_{a}\wh{\wt{t}}_{b}-s_a\wh{\wt{w}}_b=}{} +(p+q-\alpha_3)v_a\wh{\wt{w}}_b-\frac{p_a}{q-a}\wh{\wt{S}}_{a,b}+\frac{p_b}{q-a}\wh{S}_{a,b}.\label{eq:va 2}
\end{gather}
Then, noticing that
\begin{gather*}
\frac{\frac{p_b\wt{v}_a \wh{w}_b}{p-a}-\frac{p_b\wh{v}_a\wh{w}_b}{q-a}}
{(p-b)\wt{w}_b-(q-b)\wh{w}_b}w_b=\frac{p_{b}\wh{w}_{b} \frac{1}{(p-a)(q-a)}
 \big[-(p-a)\wh{v}_{a}+(q-a)\wt{v}_{a}\big]}{(p-b)\frac{\wt{w}_b}{w_b}-(q-b)\frac{\wh{w}_b}{w_b}},
\end{gather*}
and from \eqref{eq:svurels_b} and \eqref{eq:twurels_b}
we can express $v_a$ terms using $w_b$ and its shifts, and then we get
\begin{gather*}
\frac{\frac{p_b\wt{v}_a \wh{w}_b}{p-a}-\frac{p_b\wh{v}_a\wh{w}_b}{q-a}}
{(p-b)\wt{w}_b-(q-b)\wh{w}_b}w_b=-\frac{p_{b}}{(p-a)(q-a)}\wh{w}_b \wh{\wt{v}}_a.
\end{gather*}
The right-hand side can be expressed as
the hat shift of the first part of~\eqref{eq:SSrels_ab}, i.e.,
\begin{gather*}
(p-a)\wh{S}_{a,b}-(p-b)\wh{\wt{S}}_{a,b}= \wh{\wt{v}}_a \wh{w}_b.
\end{gather*}
Thus, we have
\begin{gather}
\frac{\frac{p_b\wt{v}_a \wh{w}_b}{p-a}-\frac{p_b\wh{v}_a\wh{w}_b}{q-a}}
{(p-b)\wt{w}_b-(q-b)\wh{w}_b}w_b
 =-\frac{p_{b}}{q-a}\wh{S}_{a,b} +\frac{p_b(p-b)}{(p-a)(q-a)}\wh{\wt{S}}_{a,b}.
\label{eq:va 7}
\end{gather}
Then using the relations \eqref{eq:va 2}, \eqref{eq:va 7} and making some algebras, we get
\begin{gather}
 v_{a}\wh{\wt{t}}_{b}-s_a\wh{\wt{w}}_b=
\frac{\frac{q_b\wh{v}_a \wt{w}_b}{q-a}-\frac{p_b\wt{v}_a \wh{w}_b}{p-a}}
{(p-b)\wt{w}_b-(q-b)\wh{w}_b} w_b+(p+q-\alpha_3)v_a\wh{\wt{w}}_b\nonumber\\
\hphantom{v_{a}\wh{\wt{t}}_{b}-s_a\wh{\wt{w}}_b=}{}
+\wh{\wt{S}}_{a,b}
\left(\frac{-p_a}{q-a}+\frac{p_b (p-b)}{(q-a)(p-a)}\right).\label{eq:va 8}
\end{gather}
Since
\begin{gather*}
p_{a}=\frac{G_{4}(-p,-a)}{p-a},\qquad p_{b}=\frac{G_{4}(-p,-b)}{p-b},
\end{gather*}
and
\begin{eqnarray*}
\frac{-p_a}{q-a}+\frac{p_b (p-b)}{(q-a)(p-a)}=\frac{1}{(p-a)(q-a)}\big(G_{4}(-a,-b)\big),
\end{eqnarray*}
\eqref{eq:va 8} gives rise to \eqref{eq:vwSrels-1}.

In view of the equations \eqref{eq:strels_a} and \eqref{eq:strels_b} and their hat-$q$ versions,
we delete variable $u_0$ and obtain
\begin{subequations}
\label{eq:sv-tw}
\begin{gather}
\label{eq:sveq-1}
 \wt{s}_a/\wt{v}_a-\wh{s}_a/\wh{v}_a
=p-q-v_a\big((p-a)/\wt{v}_a-(q-a)/\wh{v}_a\big), \\
\label{eq:tweq-1}
 \wt{t}_b/\wt{w}_b-\wh{t}_b/\wh{w}_b
=p-q-\wh{\wt{w}}_b\big((p-b)/\wh{w}_b-(q-b)/\wt{w}_b\big).
\end{gather}
\end{subequations}
The system composed by \eqref{eq:SSrels_ab}, \eqref{eq:vwSrels-1} and \eqref{eq:sv-tw} can be viewed as a closed-form
for elements $s_a$, $v_a$, $t_b$, $w_b$ and $S_{a,b}$, i.e.,
\begin{subequations}
\label{eq:C3}
\begin{gather}
 (p-a)S_{a,b}-(p-b)\wt{S}_{a,b}=\wt{v}_aw_b, \qquad (q-a)S_{a,b}-(q-b)\wh{S}_{a,b}=\wh{v}_a w_b,
\label{3.35a}\\
 \wt{s}_a/\wt{v}_a-\wh{s}_a/\wh{v}_a
=p-q-v_a\big((p-a)/\wt{v}_a-(q-a)/\wh{v}_a\big), \label{3.35b}\\
 \wt{t}_b/\wt{w}_b-\wh{t}_b/\wh{w}_b
=p-q-\wh{\wt{w}}_b\big((p-b)/\wh{w}_b-(q-b)/\wt{w}_b\big), \label{3.35c}\\
 v_{a}\wh{\wt{t}}_{b}-s_a\wh{\wt{w}}_{b}=-w_b\frac{\frac{p_b}{p-a}\wh{w}_b\wt{v}_a
-\frac{q_b}{q-a}\wt{w}_b\wh{v}_a}{(p-b)\wt{w}_b-(q-b)\wh{w}_b} \nonumber\\
 \hphantom{v_{a}\wh{\wt{t}}_{b}-s_a\wh{\wt{w}}_{b}=}{}+(p+q-\alpha_3)v_a\wh{\wt{w}}_{b}+\frac{G_4(-a,-b)}{(p-a)(q-a)}\wh{\wt{S}}_{a,b}.\label{C3-5}
\end{gather}
\end{subequations}
We note that the equation \eqref{C3-5} can alternatively be replaced by the following equation
\begin{gather}
\label{eq:vwSrels-2}
v_{a}\wh{\wt{t}}_{b}-s_a\wh{\wt{w}}_{b}=-w_b\frac{\frac{p_a}{p-b}\wh{w}_b\wt{v}_a
-\frac{q_a}{q-b}\wt{w}_b\wh{v}_a}{(p-b)\wt{w}_b-(q-b)\wh{w}_b}
+(p+q-\alpha_3)v_a\wh{\wt{w}}_{b}+\frac{G_4(-a,-b)}{(p-b)(q-b)}S_{a,b},
\end{gather}
which can be derived in a similar way to \eqref{eq:vwSrels-1}.
Therefore, we may have another closed form composed by
\eqref{3.35a}, \eqref{3.35b}, \eqref{3.35c} and \eqref{eq:vwSrels-2}.

\section{Lattice GD-4 type equations}\label{sec-4}

We have got quadrilateral closed forms \eqref{eq:A2-1}, \eqref{eq:A2-2}, \eqref{eq:B2} and \eqref{eq:C3}.
All the elements, $v_a$, $s_a$, $r_a$, $u_0$, etc., are well defined through \eqref{eq:objs},
which means these closed forms (if they are considered as equations) allow multi-soliton solutions.\footnote{One can introduce a particular measure $d \lambda_j(k)$ in~\eqref{eq:bC}
and this will finally lead to an explicit expression of~$\bU$ by means of a Cauchy matrix.
Thus the objects $v_a$, $w_b$, $u_{i,j}$, etc are well defined.
One can refer to~\cite[Section~5]{ZZN} for the DBSQ case
and \cite[Section 6.4]{NSZ-CMP-2022} for the elliptic solution case.}
In an ideal multicomponent quadrilateral equation, all dependent variables should evolve under certain boundary conditions.
In this section, first, we will revisit these obtained closed forms.
We will try introducing extra equations so that these closed forms \eqref{eq:A2-1}, \eqref{eq:A2-2} and \eqref{eq:B2}
can well evolve for some given boundary conditions, and therefore they can be considered
in a more general sense as lattice equations of the GD-4 type.
After that, we will provide deformations of these equations.
Evolutions of these deformed lattice equations will be examined in Section~\ref{sec-4-3}.
We will see that these lattice equations (see \eqref{eq:A2-111}, \eqref{eq:A2-222}, \eqref{eq:B222})
are the counterparts of the
Hietarinta's (A-2), (B-2) and (C-3) equations of the DBSQ type \cite{H,ZZN}.

\subsection{Closed forms: revisited}\label{sec-4-1}

The purpose of this part is to look for extra equations
to compensate for the previously obtained closed forms
so that the new systems can evolve under various boundary conditions.

In the closed form \eqref{eq:A2-1}, $\wh{\wt{r}}_a$ and $u_{0,1}$ are missed.
In the next we look for an equation that involves with $\wh{\wt{r}}_a$ and $u_{0,1}$.
For this sake, take $\wh{\phantom{a}}$-shift of \eqref{eq:zrrels_a} and using
\eqref{eq:u00-eq1} to delete $\wh{u}_{0,1}$, we get
\begin{subequations}
\begin{gather*}
\wh{\wt{r}}_a=p\wh{\wt{s}}_a-(p-a)\wh{s}_a+\wh{\wt{v}}_a\big(q\wh{u}_0-qu_0-u_{1,0}+\wh{u}_0 u_0\big).
\end{gather*}
Replacing the shift $\wt{\phantom{a}}$ by the shift $\wh{\phantom{a}}$ whilst replacing
the parameter $p$ by $q$, we have another equation on $\wh{\wt{r}}_a$:
\begin{gather*}
\wh{\wt{r}}_a=q\wh{\wt{s}}_a-(q-a)\wt{s}_a+\wh{\wt{v}}_a\big(p\wt{u}_0-pu_0-u_{1,0}+\wt{u}_0 u_0\big).
\end{gather*}
\end{subequations}
Adding them together yields
\begin{gather*}
 2\wh{\wt{r}}_a=(p+q)\wh{\wt{s}}_a-(p-a)\wh{s}_a -(q-a)\wt{s}_a \nn \\
\hphantom{2\wh{\wt{r}}_a=}{} +\wh{\wt{v}}_a\big(p\wt{u}_0+q\wh{u}_0-(p+q-\wt{u}_0-\wh{u}_0)u_0-2u_{1,0}\big).
\end{gather*}
This equation together with the closed form \eqref{eq:A2-1} can make up a new set of lattice equations, i.e.,\looseness=-1
\begin{subequations}\label{eq:A2-11}
\begin{gather}\label{eq:bA2-a1}
 \wt{s}_a=(p+u_0)\wt{v}_a-(p-a) v_a, \qquad \wh{s}_a=(q+u_0)\wh{v}_a-(q-a) v_a,\\
\label{eq:bA2-b1}
 \big(\wt{r}_a-p\wt{s}_a\big)/\wt{v}_a-\big(\wh{r}_a-q\wh{s}_a\big)/\wh{v}_a=-(p-a)s_a/\wt{v}_a+(q-a)s_a/\wh{v}_a, \\
 2\wh{\wt{r}}_a=(p+q)\wh{\wt{s}}_a-(p-a)\wh{s}_a -(q-a)\wt{s}_a \nn \\
 \hphantom{2\wh{\wt{r}}_a=}{}+\wh{\wt{v}}_a \big(p\wt{u}_0+q\wh{u}_0-\big(p+q-\wt{u}_0-\wh{u}_0\big)u_0-2u_{1,0}\big),\label{eq:bA2-c1} \\
\label{eq:bA2-re-c}
 \wh{u}_{1,0}-\wt{u}_{1,0}=\big(p-q+\wh{u}_0-\wt{u}_0\big)\wh{\wt{u}}_0-p\wh{u}_0+q\wt{u}_0, \\
 \big[H(p,q)+\wh{\wt{u}}_{1,0}-(p+q-\alpha_3)\wh{\wt{u}}_{0}
+ \big((p+q-\alpha_3-\wh{\wt{u}}_{0})s_a+r_a\big)/v_a\big] \nn \\
 \qquad {}\times \big(p-q+\wh{u}_0-\wt{u}_0\big)=(p_a\wt{v}_a-q_a\wh{v}_a)/v_a,\label{eq:bA2-d1}
\end{gather}
\end{subequations}
which evolves when initial values are given on the staircase (see Section~\ref{sec-4-3} and Figure~\ref{F1}).
In a~quite similar fashion, one can elaborate \eqref{eq:A2-2}, and we get
\begin{subequations}
\label{eq:A2-22}
\begin{gather}
\label{eq:bA2-c-re-a}
 t_b =(p-b)\wt{w}_b-\big(p-\wt{u}_0\big)w_b, \quad t_b= (q-b)\wh{w}_b-\big(q-\wh{u}_0\big)w_b, \\
\label{eq:bA2-c-re-b}
 \big(\wh{z}_b+p\wh{t}_b\big)/\wh{w}_b-\big(\wt{z}_b+q\wt{t}_b\big)/\wt{w}_b
=(p-b)\wh{\wt{t}}_b/\wh{w}_b-(q-b)\wh{\wt{t}}_b/\wt{w}_b, \\
 \wh{u}_{0,1}-\wt{u}_{0,1}=\big(p-q+\wh{u}_0-\wt{u}_0\big)u_0-p\wt{u}_0+q\wh{u}_0, \\
 2z_b = (p-b) \wt{t}_b+(q-b) \wh{t}_b-(p+q) t_b \nn \\
 \hphantom{2z_b =}{}+w_b \big(\big(p+q+\wt{u}_{0}+\wh{u}_{0}\big)\wh{\wt{u}}_0
-2\wh{\wt{u}}_{0,1}-q\wt{u}_0-p\wh{u}_0\big), \label{eq:bA2-c-re-c}\\
\big[H(p,q)+u_{0,1}+(p+q-\alpha_3)u_{0}
- \big((p+q-\alpha_3+u_{0})\wh{\wt{t}}_b-\wh{\wt{z}}_b\big)/\wh{\wt{w}}_b\big] \nn \\
 \qquad {} \times \big(p-q+\wh{u}_0-\wt{u}_0\big)=\big(p_b\wh{w}_b-q_b\wt{w}_b\big)/\wh{\wt{w}}_b,\label{eq:bA2-c-re-d}
\end{gather}
\end{subequations}
where \eqref{eq:bA2-c-re-c} is the extra equation we added.
This system can evolve with initial values that are given on the staircase.

For the closed form \eqref{eq:B2}, we need to have an extra equation involved with
$u_{0,2}$ and $\wh{\wt{u}}_{0,2}$.
To achieve that,
taking
$\wh{\phantom{a}}$-shift of the first equation in \eqref{eq:u01-1} and making use of
the second equation in \eqref{eq:u10-1}, we have
\begin{subequations}
\begin{gather}
\label{eq:u11-can_c}
\wh{\wt{u}}_{0,2}=p\wh{\wt{u}}_{0,1}-\big(p-\wh{\wt{u}}_{0}\big)\wh{u}_{0,1}
-(q+u_0)\wh{u}_{1,0}+qu_{1,0}+u_{2,0}.
\end{gather}
Similarly, we also have
\begin{gather}
\label{eq:u11-can_d}
\wh{\wt{u}}_{0,2}=q\wh{\wt{u}}_{0,1}-\big(q-\wh{\wt{u}}_{0}\big)\wt{u}_{0,1}-(p+u_0)\wt{u}_{1,0}+pu_{1,0}+u_{2,0}.
\end{gather}
\end{subequations}
Adding \eqref{eq:u11-can_c} to \eqref{eq:u11-can_d} yields
\begin{gather}
 2\wh{\wt{u}}_{0,2}=(p+q)\big(u_{1,0}+\wh{\wt{u}}_{0,1}\big)-\big(p-\wh{\wt{u}}_{0}\big)\wh{u}_{0,1}
-\big(q-\wh{\wt{u}}_{0}\big)\wt{u}_{0,1} \nn \\
 \qquad {}-(p+u_0)\wt{u}_{1,0}-(q+u_0)\wh{u}_{1,0}+2u_{2,0}.\label{eq:u11-can_cd}
\end{gather}
Inserting this equation into \eqref{eq:B2} leads to lattice equations
\begin{subequations}
\label{eq:B22}
\begin{gather}
\label{eq:u00-eq1ab-re}
 p\wt{u}_{0}-\wt{u}_{0,1}=pu_{0}+u_{1,0}-\wt{u}_{0} u_{0}, \qquad
q\wh{u}_{0}-\wh{u}_{0,1}=qu_{0}+u_{1,0}-\wh{u}_{0} u_{0}, \\
\label{eq:u11-can_a-re}
 \wh{u}_{2,0}-\wt{u}_{2,0}=-p\wh{u}_{1,0}+q\wt{u}_{1,0}+\big(p-q+\wh{u}_0-\wt{u}_0\big)\wh{\wt{u}}_{1,0}, \\
 \wh{u}_{0,2}-\wt{u}_{0,2}=-p\wt{u}_{0,1}+q\wh{u}_{0,1}+\big(p-q+\wh{u}_0-\wt{u}_0\big)u_{0,1},\\
 2\wh{\wt{u}}_{0,2}=(p+q)\big(u_{1,0}+\wh{\wt{u}}_{0,1}\big)-\big(p-\wh{\wt{u}}_{0}\big)\wh{u}_{0,1}
-\big(q-\wh{\wt{u}}_{0}\big)\wt{u}_{0,1} \nn \\
\hphantom{2\wh{\wt{u}}_{0,2}=} {} -(p+u_0)\wt{u}_{1,0}-(q+u_0)\wh{u}_{1,0}+2u_{2,0}, \label{eq:u11-can_b-re}\\
 G_4(-p,-q)/\big(p-q+\wh{u}_0-\wt{u}_0\big)=G_4(-p,-q)/(p-q)+H(p,q)\big(u_0-\wh{\wt{u}}_0\big) \nn \\
 \qquad {}
-(p+q-\alpha_3)\big(u_0\wh{\wt{u}}_0-\wh{\wt{u}}_{1,0}+u_{0,1}\big)
+u_0\wh{\wt{u}}_{1,0}-\wh{\wt{u}}_0u_{0,1}-\wh{\wt{u}}_{2,0}+u_{0,2},\label{eq:BSQ-eq3-re}
\end{gather}
\end{subequations}
which can evolve with initial values that are given on the staircase.

With regard to the closed form \eqref{eq:C3}, unfortunately,
we did not find such extra equations so that \eqref{eq:C3}
allows more choices for boundary conditions.

\subsection{Deformation of the lattice GD-4 type equations}\label{sec-4-2}

In~\cite{H}, Hietarinta obtained DBSQ type equations that are quadrilateral and multidimensionally consistent.
Those equations are named (A-2), (B-2) and (C-3).
Later, it is shown that Hietarinta's equations can be obtained as deformations of the extended DBSQ equations derived
from the DL approach \cite{ZZN}.
In the following, we introduce Hietarinta's form of the lattice GD-4 equations \eqref{eq:A2-11}, \eqref{eq:A2-22},
\eqref{eq:B22} and \eqref{eq:C3}.
In terms of naming conventions,
we will call \eqref{eq:A2-11} the GD-4 (A-2) equation, \eqref{eq:B22} the GD-4 (B-2) equation, and \eqref{eq:C3} the GD-4 (C-3) equation,
as they are the lattice GD-4 counterparts of the DBSQ (A-2), (B-2) and (C-3) equations. Equation~\eqref{eq:A2-22} also belongs to the (A-2) type as it is connected to \eqref{eq:A2-11} by reflection transformations.
Here, by ``deformation'' we mean to convert the equations obtained from the DL approach to
Hietarinta's form using point transformations.

{\bf GD-4 (A-2):} Through the point transformation
\begin{subequations}\label{eq:xyz-A2}
\begin{gather}
v_a=x/x_{a}, \qquad u_0=z-z_0, \qquad s_a=(y-v_ay_a)/x_a, \\
 u_{1,0}=\xi-z_0u_0-\xi_0, \qquad r_a =(\eta-z_0y+\xi_0x)/x_a,
\end{gather}
\end{subequations}
where
\begin{gather*}
 x_a=(p-a)^{-n}(q-a)^{-m}c_0, \qquad z_0=-pn-qm-c_1, \qquad y_a=x_az_0, \\
 \xi_0= \big( (np+mq+c_1)^2+\big(np^2+mq^2+c_2\big)\big)/2+c_3,
\end{gather*}
and $c_i$ are constants,
the lattice equation set \eqref{eq:A2-11} can be transformed into
\begin{subequations}
\label{eq:A2-111}
\begin{gather}
\label{eq:A2-ex_a}
\wt{y}=z\wt{x}-x, \qquad \wh{y}=z\wh{x}-x, \\
\label{eq:A2-ex_b}
\wt{\eta}\wh{x}-\wh{\eta}\wt{x}=y(\wt{x}-\wh{x}), \\
\label{eq:A2-ex_bb}
\wh{\xi}-\wt{\xi}=(\wh{z}-\wt{z})\wh{\wt{z}}, \\
\label{eq:A2-ex_eta-th}
\wh{\wt{\eta}}=\frac{(\wt{y}\wh{z}-\wh{y}\wt{z})z-\xi(\wt{y}-\wh{y})}{z(\wt{z}-\wh{z})},\\
\label{eq:A2-ex_c}
 \eta=-x\wh{\wt{\xi}}+y\wh{\wt{z}}+\alpha_3\big(y-x\wh{\wt{z}}\big)-\alpha_2x+
\frac{G_4(-p,-a)\wt{x} -G_4(-q,-a)\wh{x}}{\wh{z}-\wt{z}}.
\end{gather}
\end{subequations}
Note that the current equation \eqref{eq:A2-ex_eta-th} has been modified.
In fact, in light of the transformation~\eqref{eq:xyz-A2},
equation \eqref{eq:bA2-c1} gives rise to
\begin{equation}\label{x12}
2\wh{\wt{\eta}}=\wh{\wt{x}}\big(z(\wt{z}+\wh{z})-2\xi\big)-(\wt{y}+\wh{y}).
\end{equation}
Making use of \eqref{eq:A2-ex_a}, we have
\[\wh{\wt{x}}=\frac{\wt{y}-\wh{y}}{z(\wt{z}-\wh{z})},\]
substituting which into \eqref{x12} and then we get \eqref{eq:A2-ex_eta-th}.
We call this set of equations \eqref{eq:A2-111} the \mbox{GD-4} (A-2) equation.
Note that compared with~\eqref{x12}, the form~\eqref{eq:A2-ex_eta-th} has advantage in generating Lax pair.
Analogously, by point transformation
\begin{gather*}
 w_b=x/x_b, \qquad u_0=z-z_0, \qquad t_b=(y-w_by_b)/x_b, \\
 u_{0,1}=\xi-z_0u_0-\xi_0, \qquad z_b=(\eta-z_0y+x\xi_0)/x_b,
\end{gather*}
with
\begin{gather*}
 x_b =(-p+b)^{n}(-q+b)^{m}c_0, \qquad z_0=-pn-qm-c_1, \qquad y_b=x_bz_0,\\
 \xi_0 = \big( (np+mq+c_1)^2-\big(np^2+mq^2+c_2\big)\big)/2-c_3,
\end{gather*}
and constants $c_i$,
the system \eqref{eq:A2-22} gives rise to
\begin{subequations}
\label{eq:A2-222}
\begin{gather}
\label{eq:A2-ex-N4-1 a}
 y=\wt{z} x-\wt{x}, \qquad y=\wh{z}x-\wh{x}, \\
\label{eq:A2-ex-N4-1 b}
 \wt{\eta}\wh{x}-\wh{\eta}\wt{x}=\wh{\wt{y}}(\wt{x}-\wh{x}), \\
\label{eq:A2-ex-N4-1 bb}
 \wh{\xi}-\wt{\xi}=(\wh{z}-\wt{z})z, \\
\label{eq:alt-A2-ex_eta}
 \wh{\wt{\xi}}=\frac{-\eta(\wh{x}-\wt{x})+y(\wh{y}-\wt{y})+ \wt{x}\wh{y}- \wh{x}\wt{y}}{x(\wh{x}-\wt{x})},\\
\label{eq:A2-ex-N4-1 c}
\wh{\wt{\eta}}=-\wh{\wt{x}}\xi+\wh{\wt{y}}z-\alpha_3\big(\wh{\wt{y}}-\wh{\wt{x}}z\big)-\alpha_2 \wh{\wt{x}}
-\frac{G_4(-p,-b)\wh{x}-G_4(-q,-b)\wt{x}}{\wh{z}-\wt{z}}.
\end{gather}
\end{subequations}
Here we already modified \eqref{eq:alt-A2-ex_eta}.
In fact, \eqref{eq:bA2-c-re-c} gives rise to
\begin{equation}\label{eta}
2\eta=-x\big(2\wh{\wt{\xi}}-(\wt{z}+\wh{z})\wh{\wt{z}}\big)-(\wt{y}+\wh{y}).
\end{equation}
From \eqref{eq:A2-ex-N4-1 a}, we have
\[\wh{\wt{z}}= \frac{\wh{y}-\wt{y}}{\wh{x}-\wt{x}},\qquad
x(\wt{z}+\wh{x})= 2y +\wt{x}+\wh{x},\]
using which we eliminate $z$ variable from \eqref{eta} yields \eqref{eq:alt-A2-ex_eta}.
Equations \eqref{eq:A2-222} compose an alternative form of the GD-4 (A-2) equation.
It is related to \eqref{eq:A2-111} by reflection transformations
\begin{gather*}
p\rightarrow -p, \!\!\!\qquad q\rightarrow -q, \!\!\!\qquad n\rightarrow -n, \!\!\!\qquad m\rightarrow -m, \!\!\!\qquad
\alpha_3 \rightarrow -\alpha_3, \!\!\!\qquad \alpha_1 \rightarrow -\alpha_1, \!\!\!\qquad b \rightarrow -a.
\end{gather*}
Besides, both equation \eqref{eq:A2-111} and \eqref{eq:A2-222} allow a symmetry
$(\,\wt{~},p) \leftrightarrow (\,\wh{~},q)$.

{\bf GD-4 (B-2):} We adopt the point transformation
\begin{gather*}
 u_0=x-x_{0}, \qquad u_{1,0}=y-x_0u_0-y_0, \qquad u_{0,1}=z-x_0u_0-z_0, \\
 u_{2,0}=\xi-x_0y+zz_0+\mu, \qquad u_{0,2}=\eta-x_0z+zy_0+\nu,
\end{gather*}
where
\begin{gather*}
 x_0=-pn-qm-c_1, \\
 y_0= \big( (np+mq+c_1)^2+\big(np^2+mq^2+c_2\big)\big)/2+c_4,\\
 z_0= \big( (np+mq+c_1)^2-\big(np^2+mq^2+c_2\big)\big)/2-c_4,\\
 \mu_0= \big(np^{3}+mq^{3}+c_3\big)/3,\\
 \mu=-\frac{1}{6}x^{3}_{0}+\frac{1}{2}(y_{0}-z_{0})x_{0} +\frac{1}{3}\mu_0+c_{5},\\
 \nu=-\frac{1}{6}x^{3}_{0}-\frac{1}{2}(y_{0}-z_{0})x_{0} +\frac{1}{3}\mu_0+c_{5},
\end{gather*}
and $c_{i}$ are constants.
Then from \eqref{eq:B22} we have the GD-4 (B-2) equation
\begin{subequations}
\label{eq:B222}
\begin{gather}
\label{eq:BSQ-ex_a}
\wt{z}=x\wt{x}-y, \qquad \wh{z}=x\wh{x}-y, \\
\label{eq:BSQ-ex_b}
 \wh{\xi}-\wt{\xi}=(\wh{x}-\wt{x})\wh{\wt{y}}, \\
\label{eq:BSQ-ex_bb}
 \wh{\eta}-\wt{\eta}=(\wh{x}-\wt{x})z, \\
\label{eq:B2-ex_eta-th}
 \wh{\wt{\eta}}=\xi + \frac{(\wh{x}\wt{y}-\wt{x}\wh{y})x-y(\wt{y}-\wh{y})}{\wt{x}-\wh{x}},\\
\label{eq:BSQ-ex_c}
 \eta=\wh{\wt{\xi}}-\wh{\wt{y}}x+\wh{\wt{x}}z+\alpha_3\big(\wh{\wt{y}}-x\wh{\wt{x}}+z\big)
+\alpha_2\big(\wh{\wt{x}}-x\big)+\frac{G_4(-p,-q)}{\wh{x}-\wt{x}}.
\end{gather}
\end{subequations}
Again, we have modified \eqref{eq:B2-ex_eta-th}.
Under the transformation \eqref{eq:u11-can_b-re} gives rise to
\begin{equation}\label{eta12}
2\wh{\wt{\eta}}=2\xi-x(\wh{y}+\wt{y})+(\wh{z}+\wt{z})\wh{\wt{x}}.
\end{equation}
From \eqref{eq:BSQ-ex_a}, we can have
\[\wh{\wt{x}}= \frac{\wt{y}-\wh{y}}{\wt{x}-\wh{x}},\qquad \wt{z}+\wh{z}= (\wt{x}+\wh{x})x-2y.\]
Substituting them into \eqref{eta12} yields \eqref{eq:B2-ex_eta-th}.
Note that \eqref{eq:B222} allows a symmetry
$(\,\wt{~},p) \leftrightarrow (\,\wh{~},q)$.

{\bf GD-4 (C-3):} Inserting point transformation
\begin{subequations}
\label{eq:MSBSQ-tran}
\begin{align}
& S_{a,b}= \big( (p-a)/(p-b)\big)^n \big( (q-a)/(q-b)\big)^m x, \\
& v_a=(p-a)^n(q-a)^m y, \quad w_b=(p-b)^{-n}(q-b)^{-m} z, \\
& s_a=(p-a)^n(q-a)^m(y_1-z_0y), \\
& t_b=(p-b)^{-n}(q-b)^{-m}(z_1-z_0z),
\end{align}
\end{subequations}
into lattice equation \eqref{eq:C3}, where $z_0= -pn-qm-c_{1}$,
we get the GD-4 (C-3) equation
\begin{subequations}
\label{eq:xyz-MSBSQ}
\begin{gather}
\label{eq:xyz-MSBSQ-a}
 x-\wt{x}=\wt{y}z, \qquad x-\wh{x}=\wh{y}z, \\
\label{eq:xyz-MSBSQ-b}
 (\wt{y}_1+y)\wh{y}=(\wh{y}_1+y)\wt{y}, \quad \big(\wh{z}_1-\wh{\wt{z}}\big)\wt{z}
=\big(\wt{z}_1-\wh{\wt{z}}\big)\wh{z}, \\
\label{eq:xyz-MSBSQ-c}
 y \wh{\wt{z}}_1-y_1\wh{\wt{z}}=
z\frac{G_4(-p,-b)\wh{z}\wt{y}-G_4(-q,-b)\wt{z}\wh{y}}{\wh{z}-\wt{z}}-\alpha_3y\wh{\wt{z}}
+G_4(-a,-b)\wh{\wt{x}}.
\end{gather}
\end{subequations}
The alternative GD-4 (C-3) equation is
\begin{subequations}
\label{eq:xyz-MSBSQ-1}
\begin{gather}
\label{eq:xyz-MSBSQ-1a}
 x-\wt{x}=\wt{y}z, \qquad x-\wh{x}=\wh{y}z, \\
\label{eq:xyz-MSBSQ-1b}
 (\wt{y}_1+y)\wh{y}=(\wh{y}_1+y)\wt{y},
\qquad \big(\wh{z}_1-\wh{\wt{z}}\big)\wt{z}=\big(\wt{z}_1-\wh{\wt{z}}\big)\wh{z}, \\
\label{eq:xyz-MSBSQ-1c}
 y\wh{\wt{z}}_1-y_1 \wh{\wt{z}}=z\frac{G_4(-p,-a)\wh{z}\wt{y}-G_4(-q,-a)\wt{z}\wh{y}}{\wh{z}-\wt{z}}
-\alpha_3y\wh{\wt{z}}+G_4(-a,-b)x,
\end{gather}
\end{subequations}
where the last equation is the deformation of \eqref{eq:vwSrels-2} in light of the transformation \eqref{eq:MSBSQ-tran}.
Equations \eqref{eq:xyz-MSBSQ} and
\eqref{eq:xyz-MSBSQ-1} are related by the reversal symmetry transformation
\begin{gather*}
n\rightarrow -n, \qquad m\rightarrow -m, \qquad y\rightarrow z, \qquad z\rightarrow -y, \qquad y_1\rightarrow z_1, \qquad z_1\rightarrow -y_1,
 \qquad a \leftrightarrow b.
 \end{gather*}
Both equation \eqref{eq:xyz-MSBSQ} and \eqref{eq:xyz-MSBSQ-1} allow a symmetry
$(\,\wt{~},p) \leftrightarrow (\,\wh{~},q)$.
Besides, \eqref{eq:xyz-MSBSQ-c} and \eqref{eq:xyz-MSBSQ-1c} are also connected through
\begin{align*}
z(\wh{z}\wt{y}-\wt{z}\wh{y})=\big(\wh{\wt{x}}-x\big)(\wt{z}-\wh{z}),
\end{align*}
which holds in light of \eqref{eq:xyz-MSBSQ-a}.
In addition, we note that, when $a=b$ in \eqref{eq:xyz-MSBSQ-1c} and we eliminate~$x$ from \eqref{eq:xyz-MSBSQ-1a},
we can reduce the GD-4 (C-3) equation \eqref{eq:xyz-MSBSQ-1} to the following four-component form
\begin{gather*}
 \wh{\wt y}(\wt z-\wh z)=-z(\wt y-\wh y), \qquad
 (\wt{y}_1+y)\wh{y}=(\wh{y}_1+y)\wt{y},
\qquad \big(\wh{z}_1-\wh{\wt{z}}\big)\wt{z}=\big(\wt{z}_1-\wh{\wt{z}}\big)\wh{z}, \\
 y\wh{\wt{z}}_1-y_1 \wh{\wt{z}}=z\frac{G_4(-p,-a)\wh{z}\wt{y}-G_4(-q,-a)\wt{z}\wh{y}}{\wh{z}-\wt{z}}
-\alpha_3y\wh{\wt{z}}.
\end{gather*}

{\bf GD-4 (C-4):}
In \cite{H}, Hietarinta also obtained a (C-4) equation of the DBSQ type (see~\eqref{eq:C4-N=3}),
which was later shown to be another deformation of the (C-3) equation.
In the following, we show a similar deformation and present the GD-4 (C-4) equation.

First, we observe that
equations \eqref{eq:xyz-MSBSQ} and \eqref{eq:xyz-MSBSQ-1} share the same solution
through \eqref{eq:MSBSQ-tran}.
Thus, from \eqref{eq:xyz-MSBSQ} and \eqref{eq:xyz-MSBSQ-1}, we have
\begin{subequations}
\label{eq:C3-s}
\begin{gather}
\label{eq:C3-s1}
 x-\wt{x}=\wt{y}z, \qquad x-\wh{x}=\wh{y}z, \\
\label{eq:C3-s2}
 (\wt{y}_1+y)\wh{y}=(\wh{y}_1+y)\wt{y}, \qquad \big(\wh{z}_1-\wh{\wt{z}}_1\big)\wt{z}
=\big(\wt{z}_1-\wh{\wt{z}}_1\big)\wh{z}, \\
\label{eq:C3-s3}
 y \wh{\wt{z}}_1-y_1 \wh{\wt{z}}=z\frac{P_{a,b}\wh{z}\wt{y}-Q_{a,b}\wt{z}\wh{y}}{\wh{z}-\wt{z}}
-\alpha_3y\wh{\wt{z}}+G_{a,b}\big(\wh{\wt{x}}+x\big),
\end{gather}
\end{subequations}
where
\begin{gather*}
 P_{a,b}= \big(G_4(-p,-b)+G_4(-p,-a)\big)/2, \\
 Q_{a,b}= \big(G_4(-q,-b)+G_4(-q,-a)\big)/2, \qquad
 G_{a,b}=G_4(-a,-b)/2.
\end{gather*}
Then, consider the following transformation
\begin{gather}
x=\frac{x_2-G_{a,b}}{2G_{a,b}(x_2+G_{a,b})}, \qquad y=\frac{y_2}{x_2+G_{a,b}},
\qquad z=\frac{z_2}{x_2+G_{a,b}},\nonumber\\
 y_1=\frac{y_3}{x_2+G_{a,b}}, \qquad z_1=\frac{z_3}{x_2+G_{a,b}}.\label{eq:C3-C4}
\end{gather}
Imposing this transformation on~\eqref{eq:C3-s1} and~\eqref{eq:C3-s2} gives rise to
\begin{gather*}
 x_2-\wt{x}_2=\wt{y}_2z_2, \qquad x_2-\wh{x}_2=\wh{y}_2z_2, \\
 (\wt{y}_3+y_2)\wh{y}_2=(\wh{y}_3+y_2)\wt{y}_2,
\qquad \big(\wh{z}_3-\wh{\wt{z}}_2\big)\wt{z}_2=\big(\wt{z}_3-\wh{\wt{z}}_2\big)\wh{z}_2,
\end{gather*}
which further yields
\begin{gather}
\label{eq:con-a}
\wh{\wt{x}}_2=\frac{\wh{x}_2\wt{z}_2-\wt{x}_2\wh{z}_2}{\wt{z}_2-\wh{z}_2}.
\end{gather}
Meanwhile, by the transformation \eqref{eq:C3-C4}, equation \eqref{eq:C3-s3}
multiplied by $(G_{a,b}+x_2)\big(G_{a,b}+\wh{\wt{x}}_2\big)$
is equivalent to
\begin{gather}
\label{eq:C4-cal}
y_2 \wh{\wt z}_3-y_3 \wh{\wt z}_2=\Delta-\alpha_3y_2
\wh{\wt z}_2 + x_2 \wh{\wt x}_2-G^2_{a,b},
\end{gather}
where
\begin{align*}
\Delta= \frac{z_2\big(G_{a,b}+\wh{\wt
x}_2\big)(P_{a,b}\,\wh{z}_2\wt{y}_2-Q_{a,b}\,\wt{z}_2\wh{y}_2)}{
 (\wh{z}_2\wt{x}_2-\wt{z}_2\wh{x}_2)+G_{a,b}(\wh{z}_2-\wt{z}_2)}.
\end{align*}
We substitute \eqref{eq:con-a} into $\Delta$ and rewrite \eqref{eq:C4-cal} as
\begin{align*}
y_2 \wh{\wt z}_3-y_3 \wh{\wt z}_2=z_2\,(P_{a,b}\,\wh{z}_2\wt{y}_2
-Q_{a,b}\,\wt{z}_2\wh{y}_2)/(\wh{z}_2-\wt{z}_2)-\alpha_3y_2
\wh{\wt z}_2 + x_2 \wh{\wt x}_2-G^2_{a,b}.
\end{align*}
Finally, we arrive at the GD-4 (C-4) equation
\begin{subequations}
\label{eq:C4-s}
\begin{gather}
\label{eq:C4-s1}
 x_2-\wt{x}_2= \wt{y}_2z_2, \qquad x_2-\wh{x}_2=\wh{y}_2z_2, \\
\label{eq:C4-s2}
 (\wt{y}_3+y_2)\wh{y}_2=(\wh{y}_3+y_2)\wt{y}_2,
\qquad \big(\wh{z}_3-\wh{\wt{z}}_2\big)\wt{z}_2=\big(\wt{z}_3-\wh{\wt{z}}_2\big)\wh{z}_2, \\
\label{eq:C4-s3}
 y_2 \wh{\wt z}_3-y_3 \wh{\wt z}_2=z_2\,(P_{a,b}\,\wh{z}_2\wt{y}_2
-Q_{a,b}\,\wt{z}_2\wh{y}_2)/(\wh{z}_2-\wt{z}_2)-\alpha_3y_2
\wh{\wt z}_2 + x_2 \wh{\wt x}_2-G^2_{a,b}.
\end{gather}
\end{subequations}

{\bf Four-component GD-4 (A-2)}: In order to reduce the GD-4 (A-2) equation \eqref{eq:A2-111} into
four-component form, we introduce variables
\begin{equation*}
u=\frac{y}{x},\qquad v=\frac{\eta}{x}.
\end{equation*}
Then, equation \eqref{eq:A2-111}
can be rewritten as
\begin{subequations}
\label{eq:4comp-A-21}
\begin{gather}
\label{eq:4comp-A-2 a}
 (\wh{u}-z)\big(\wh{\wt{u}}-\wh{z}\big)=(\wt{u}-z)\big(\wh{\wt{u}}-\wt{z}\big), \\
\label{eq:4comp-A-2 bc}
 \wt{v}-\wh{v}=u(\wt{u}-\wh{u}), \\
\label{eq:4comp-A-2 bb}
 \wh{\xi}-\wt{\xi}=(\wh{z}-\wt{z})\wh{\wt{z}}, \\
\label{eq:4comp-A-2 c}
 \wh{\wt{v}}=\frac{1}{z(\wt{z}-\wh{z})} \big[z \big((\wt{u}-\wh{u})\wt{z}\wh{z}
+(\wh{u}\wt{z}-\wt{u}\wh{z})\wh{\wt{u}}\big)-\xi \big(\wt{u}\wt{z}
-\wh{u}\wh{z}+(\wt{u}-\wt{u})\wh{\wt{u}}\big)\big], \\
\label{eq:4comp-A-2 d}
 (\wh{z}-\wt{z}) \big[v+\wh{\wt{\xi}}-(u-\alpha_{3})\wh{\wt{z}}-\alpha_{3} u+\alpha_{2}\big]
=\frac{G_{4}(-p,-a)}{z-\wt{u}}-\frac{G_{4}(-q,-a)}{z-\wh{u}}.
\end{gather}
\end{subequations}
We view this system as the four-component GD-4 (A-2) equation.
Similarly, by introducing variables
$w=\frac{y}{x}$, $\varpi=\frac{\eta}{x}$,
system \eqref{eq:A2-222} becomes
\begin{subequations}
\label{eq:4comp-A-22}
\begin{gather}
\label{eq:4comp-3.41a}
 \big(\wh{\wt{z}}-\wh{w}\big)(\wh{z}-w)=\big(\wh{\wt{z}}-\wt{w}\big)(\wt{z}-w), \\
\label{eq:4comp-3.41b}
 \wt{\varpi}-\wh{\varpi}=\wh{\wt{w}}(\wt{w}-\wh{w}), \\
\label{eq:4comp-3.41bb}
 \wh{\xi}-\wt{\xi}=(\wh{z}-\wt{z})z, \\
\label{eq:4comp-3.41c}
 \wh{\wt{\xi}}=\frac{-\varpi (\wh{z}-\wt{z})+w \big(\wh{w}(\wh{z}-w)-\wt{w}(\wt{z}-w)\big)
+\wh{w}(\wt{z}-w)(\wh{z}-w)-\wt{w}(\wh{z}-w)(\wt{z}-w)}{\wh{z}-\wt{z}},\!\!\!\!\! \\
\label{eq:4comp-3.41d}
 \big(\wh{\wt{\varpi}}+\xi-z\wh{\wt{w}}+\alpha_{3}(\wh{\wt{w}}-z)+\alpha_{2}\big)(\wh{z}-\wt{z})
=\frac{G_{4}(-p,-b)}{\wh{u}-\wh{\wt{z}}}-\frac{G_{4}(-q,-b)}{\wt{u}-\wh{\wt{z}}},
\end{gather}
\end{subequations}
which is viewed as the alternative four-component GD-4 (A-2) equation.

Note that both the equations in terms of DL notations \eqref{eq:objs}
and in Hietarinta's form in terms of $x$, $y$, $z$, etc can be used to express GD-4 lattice equations.
The DL approach enables us to see how these equations follow from one and the same underlying structure,
while the equations in Hietarinta's form are more convenient to investigate their evolution and
MDC property.

As for the names (A-2), (B-2) and (C-3), we have the following comments.
On one side, as we can see, the DBSQ (A-2), (B-2) and (C-3) equations in Hietarinta's form
are embedded in the corresponding GD-4 lattice equations.
On the other side, in the DBSQ case, one needs to employ more than one variables
so that the system is quadrilateral.
Different variables from a~same system can express different equations in one-component form.
For example, the DBSQ (A-2) equation (A.2) can give rise to the modified DBSQ equation
in terms of $x$ and the regular DBSQ equation in terms of either $z$ or $w=y/x$, see
\cite[Section~3.2]{HZ-2021}.
In this context, it is not correct to call the DBSQ (A-2) equation the modified DBSQ
or regular DBSQ equation.
We prefer to use (A-2), (B-2) and (C-3) in the GD-4 case.

\subsection{General description of MDC}\label{sec-5-1}

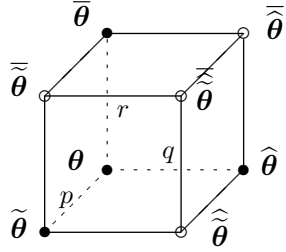
\begin{figure}[t]\centering
\setlength{\unitlength}{0.0004in} \hspace{6cm}
\begin{picture}(9482,3000)(0,0)
\put(450,1883){\circle{150}}
\put(0,1883){\makebox(0,0)[lb]{$\wb{\wt{\Tht}}$}}
\put(1275,2708){\circle*{150}} \put(825,2808){\makebox(0,0)[lb]{$\wb{\Tht}$}} \put(3075,2708){\circle{150}}
\put(3375,2633){\makebox(0,0)[lb]{$\wb{\wh{\Tht}}$}}
\put(2250,83){\circle{150}} \put(2650,-100){\makebox(0,0)[lb]{$\wh{\wt{\Tht}}$}}
\put(1275,908){\circle*{150}}
\put(750,908){\makebox(0,0)[lb]{$\Tht$}} \put(2250,1883){\circle{150}}
\put(2450,1700){\makebox(0,0)[lb]{$\wb{\wh{\wt{\Tht}}}$}}
\put(450,83){\circle*{150}} \put(0,8){\makebox(0,0)[lb]{$\wt \Tht$}}
\put(3075,908){\circle*{150}} \put(3300,833){\makebox(0,0)[lb]{$\wh \Tht$}}
\drawline(1275,2708)(3075,2708) \drawline(1275,2708)(450,1883)
\drawline(450,1883)(450,83) \drawline(3075,2708)(2250,1883)
\drawline(450,1883)(2250,1883) \drawline(3075,2633)(3075,908)
\drawline(2250,1883)(2250,83) \drawline(450,83)(2250,83)
\drawline(3075,908)(2250,83) \dashline{60.000}(1275,908)(450,83)
\dashline{60.000}(1275,908)(3075,908)
\dashline{60.000}(1275,2633)(1275,908)
\put(650,400){\makebox(0,0)[lb]{\small$p$}}
\put(2000,1000){\makebox(0,0)[lb]{\small$q$}}
\put(1400,1600){\makebox(0,0)[lb]{\small$r$}}
\end{picture}
\caption{Consistent cube of the lattice equation~\eqref{mat-F}.\label{F2}}
\end{figure}

In the past two decades, the property of multidimensional consistency (MDC)
has been effectively used in the study of discrete integrable systems.
This property allows a lattice equation (or a system) to be consistently embedded into a higher dimension
\cite{ABS-2003,BS-QG,Nij-LP,NW-2001}.
With regard to the lattice GD-4 type equations we listed in Section~\ref{sec-4-2}, they can be written as a quad equation%
\begin{gather}\label{mat-F}
\mathcal {F}\big(\Tht,\wt{\Tht},\wh{\Tht},\wh{\wt{\Tht}};p,q\big)=0,
\end{gather}
where $\mathcal{F}$ is a nonlinear vector function of the vector $\Tht$ with several components.
The key idea of the MDC property is to embed the equation consistently into a multi-dimensional
lattice by imposing copies of the same equation, albeit with different lattice parameters
in different directions.
The GD-4 type equations are symmetric between the $(p,n)$ and
$(q,m)$ coordinates of the $\mathbb{Z}^2$ lattice, and therefore
we introduce a third dimension $l$ and the
corresponding lattice index $r$ by $\Tht\rightarrow \wb{\Tht}$ to keep this symmetry.
This means that we have the same equation on all planes around the cube (see Figure~\ref{F2}). The equations
on the six faces can be written as
\begin{subequations}
\label{mat-F-abc}
\begin{alignat}{3}
\label{eq:mat-F-a}
& \mathcal {F}\big(\Tht,\wt{\Tht},\wh{\Tht},\wh{\wt{\Tht}};p,q\big)=0 \quad (\text{bottom}), \qquad&&
\mathcal {F}\big(\wb{\Tht},\wb{\wt{\Tht}},\wb{\wh{\Tht}},\wb{\wh{\wt{\Tht}}};p,q\big)=0 \quad (\text{top}),& \\
\label{eq:mat-F-b}
& \mathcal {F}\big(\Tht,\wt{\Tht},\wb{\Tht},\wb{\wt{\Tht}};p,r\big)=0 \quad (\text{left}), \qquad&&
\mathcal {F}\big(\wh{\Tht},\wh{\wt{\Tht}},\wb{\wh{\Tht}},\wb{\wh{\wt{\Tht}}};p,r\big)=0 \quad (\text{right}),& \\
\label{eq:mat-F-c}
& \mathcal {F}\big(\Tht,\wb{\Tht},\wh{\Tht},\wh{\wb{\Tht}};q,r\big)=0 \quad (\text{back}), \qquad&&
\mathcal {F}\big(\wt{\Tht},\wh{\wt{\Tht}},\wb{\wt{\Tht}},\wb{\wh{\wt{\Tht}}};q,r\big)=0 \quad (\text{front}).&
\end{alignat}
\end{subequations}
For a quadrilateral equation, e.g., \eqref{mat-F},
that can be solved
for the fourth variable given the first three,
its MDC property can be explained as follows:
In the case of the initial values given
at black dots \smash[b]{$\big(\Tht,\wt{\Tht},\wh{\Tht}, \wb{\Tht}\big)$} (see Figure~\ref{F2}),
we use the left-hand side equations in \eqref{eq:mat-F-a}, \eqref{eq:mat-F-b}, \eqref{eq:mat-F-c}
to compute \smash[b]{$\wh{\wt{\Tht}}$}, \smash[b]{$\wb{\wt{\Tht}}$}, \smash[b]{$\wh{\wb{\Tht}}$}, respectively,
and this leaves the three
right-hand side equations from which we should get the \smash[b]{$\wb{\wh{\wt{\Tht}}}$}.
The MDC property means that all three values thus obtained for~\smash[b]{$\wb{\wh{\wt{\Tht}}}$} coincide.
The MDC property for quadrilateral equations can be geometrically interpreted as a~consistency around a cube (CAC).
In light of CAC property, in principle, the coupled system
\begin{subequations}\label{mat-F-BT}
\begin{gather}
 \mathcal {F}\big(\Tht,\wt{\Tht},\wb{\Tht},\wt{\wb{\Tht}};p,r\big)=0,\\
 \mathcal {F}\big(\Tht,\wb{\Tht},\wh{\Tht},\wh{\wb{\Tht}};q,r\big)=0
\end{gather}
\end{subequations}
act as a B\"acklund transformation for the equation~\eqref{mat-F} to transform solutions
between~$\Tht$ and~$\wb\Tht$.
From the B\"acklund transformation, in principle, Lax pairs can be constructed \cite{ABS-2003,Bri,Nij-LP}.

\subsection{Evolution}\label{sec-4-3}

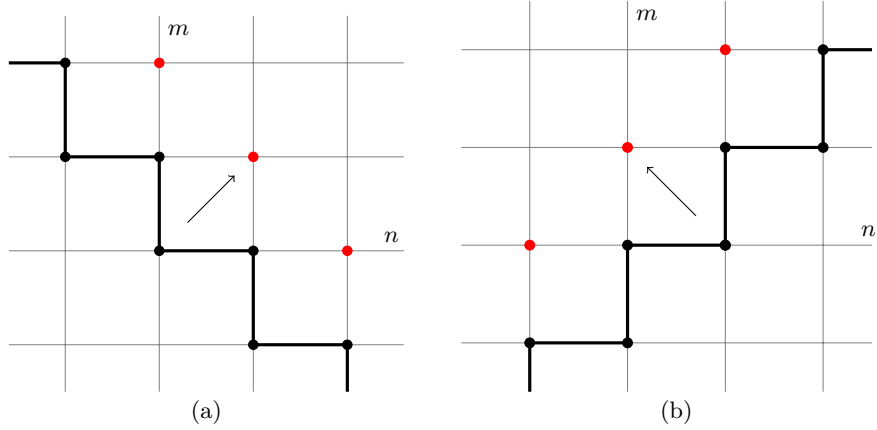
\begin{figure}[t]
{\footnotesize
\centering
\hspace*{20mm}
\begin{tikzpicture}[yscale=1.25,xscale=1.25]
\draw[step=1cm,gray,very thin] (-0.6,1.5) grid (3.6,5.5);
\draw[very thick] (-0.6,5) -- (0,5) -- (0,4) -- (1,4) -- (1,3) -- (2,3)
 -- (2,2) -- (3,2) -- (3,1.5);
\filldraw[red] (2,4) circle (0.05cm);
\filldraw[red] (3,3) circle (0.05cm);
\filldraw[red] (1,5) circle (0.05cm);
\filldraw[black] (1,3) circle (0.05cm) ;
\draw[->] (1.3,3.3) -- (1.8,3.8);
\draw[black] (1,5.5) node[below right] {$m$} ;
\draw[black] (3.3,3) node[above right] {$n$} ;
\filldraw[black] (2,3) circle (0.05cm);
\filldraw[black] (1,4) circle (0.05cm);
\filldraw[black] (0,4) circle (0.05cm);
\filldraw[black] (0,5) circle (0.05cm) ;
\filldraw[black] (2,2) circle (0.05cm);
\filldraw[black] (3,2) circle (0.05cm);
\draw[black] (1.5,1.5) node[below] {(a)} ;
\end{tikzpicture}
\hspace{0.5cm}
\begin{tikzpicture}[yscale=1.3,xscale=1.3]
\draw[step=1cm,gray,very thin] (-0.7,1.5) grid (3.5,5.5);
\draw[very thick] (3.5,5) -- (3,5) -- (3,4) -- (2,4) -- (2,3) -- (1,3)
 -- (1,2) -- (0,2) -- (0,1.5);
 \draw[->] (1.7,3.3) -- (1.2,3.8) ;
\draw[black] (1,5.5) node[below right] {$m$} ;
\draw[black] (3.3,3) node[above right] {$n$} ;
\filldraw[red] (1,4) circle (0.05cm);
\filldraw[red] (0,3) circle (0.05cm);
\filldraw[red] (2,5) circle (0.05cm);
\filldraw[black] (2,3) circle (0.05cm);
\filldraw[black] (1,3) circle (0.05cm);
\filldraw[black] (2,4) circle (0.05cm);
\filldraw[black] (3,4) circle (0.05cm);
\filldraw[black] (3,5) circle (0.05cm);
\filldraw[black] (1,2) circle (0.05cm);
\filldraw[black] (0,2) circle (0.05cm);
\draw[black] (1.5,1.5) node[below] {(b)} ;
\draw[black] (2,3) circle (0.05cm);
\end{tikzpicture}}
\caption{The initial values should be given on black dots of the staircase. One can compute the values at red dots.
This way, the equations can evolve up-right as in (a) and up-left as in (b).}\label{F1}
\end{figure}

To explore the MDC property of the GD-4 lattice equations, we analyze their evolutions.
Let us take the GD-4 (B-2) equation \eqref{eq:B222} as an example.
It is a system consisting of six equations for five variables $x$, $y$, $z$, $\xi$ and $\eta$.
In general, such a system is over-determined. We have the following remarks for such a situation:

\begin{Remark}
If we ONLY require the equations to evolve with a certain staircase initial boundary,
we will not need all 6 equations.
\end{Remark}
For convenience, we introduce $V$ to denote $V=(x, y, z, \xi, \eta)^{\rm T}$.
It turns out that, the system~\eqref{eq:B222}, excluding \eqref{eq:B2-ex_eta-th},
allows an up-left evolution for given $V$, $\wt V$, \smash[b]{$\wh{\wt V}$} as in Figure~\ref{F1}\,(b)
or a~down-right evolution for a given $V$, $\wh V$, \smash[b]{$\wh{\wt V}$}.
In addition, without~\eqref{eq:BSQ-ex_bb} the system~\eqref{eq:B222} allows an up-right evolution,
and without \eqref{eq:BSQ-ex_b} it allows a down-left evolution.
Now it is clear that we may exclude one equation from
(\ref{eq:BSQ-ex_b}, \ref{eq:BSQ-ex_bb}, \ref{eq:B2-ex_eta-th}),
and the remaining system will consist of five equations and evolve with some suitable
staircase initial values.

\begin{Remark}\label{R-4-2}
If we require the system \eqref{eq:B222} to be multidimensionally consistent,
we need ALL six equations. See Section~\ref{sec-5} and Appendix \ref{App-3}.
Similar features of evolution hold for the GD-4 (A-2) equations \eqref{eq:A2-111} and \eqref{eq:A2-222}
and for the four-component GD-4 (A-2) equations \eqref{eq:4comp-A-21} and~\eqref{eq:4comp-A-22}.
\end{Remark}

\begin{Remark}
The GD-4 (C-3) equation \eqref{eq:xyz-MSBSQ} (or \eqref{eq:xyz-MSBSQ-1})
and the GD-4 (C-4) equation \eqref{eq:C4-s} consist of 5 equations, respectively.
They can have either up-left or down-right evolution, but neither up-right nor down-left evolution.
\end{Remark}

The MDC property implies certain integrability.
Since for the GD-4 equations \eqref{eq:A2-111}, \eqref{eq:A2-222},
\eqref{eq:B222}, \eqref{eq:4comp-A-21} and \eqref{eq:4comp-A-22},
all equations are necessary to guarantee each system that they compose
to be MDC, we prefer to keep all the equations in each system.

\section{MDC property and Lax representation}\label{sec-5}

\subsection{MDC property of the GD-4 (A-2) and (B-2) equation}\label{sec-5-2}

We have examined the evolution of the GD-4 type equations in Section~\ref{sec-4-3}.
Among them, the GD-4 (A-2) equation \eqref{eq:A2-111} and its alternative form \eqref{eq:A2-222},
the GD-4 (B-2) equation \eqref{eq:B222}, four-component GD-4 (A-2) equation \eqref{eq:4comp-A-21}
 and its alternative form \eqref{eq:4comp-A-22},
allow up-right evolution with staircase initial values.
Therefore, we can check whether these equations are consistent around the cube (CAC).
Note that equations \eqref{eq:A2-ex_b}, \eqref{eq:A2-ex-N4-1 bb}, \eqref{eq:BSQ-ex_b},
\eqref{eq:4comp-A-2 bc} and \eqref{eq:4comp-3.41bb}
are not necessary to the systems where they are to get up-right evolutions.
However, in the following we will see that these equations
are really needed for their systems to be CAC.
The checking is straightforward.

{\bf{GD-4 (A-2) equation}}:
For the GD-4 (A-2) equation \eqref{eq:A2-111},
we set it to be~\eqref{mat-F}.
It is clear that from the left-hand side equations in \eqref{eq:mat-F-a}, \eqref{eq:mat-F-b}, \eqref{eq:mat-F-c},
we evaluate
the following second-order shifts
\begin{subequations}\label{A-2-CAC}
\begin{gather}
\label{A-2-CAC a}
 \wb{\wt x}=(\wt{x}-\wb{x})/(\wt{z}-\wb{z}), \qquad \wb{\wt y}=(\wt x \wb z-\wb x\wt z)/(\wt z-\wb z), \qquad
\wb{\wt z}=\big(\wt \xi-\wb \xi\big)/(\wt z-\wb z), \\
\label{A-2-CAC b}
 \wh{\wb x}=(\wb{x}-\wh{x})/(\wb{z}-\wh{z}), \qquad \wh{\wb y}=(\wb x \wh z-\wh x\wb z)/(\wb z-\wh z), \qquad
\wh{\wb z}=\big(\wb \xi-\wh \xi\big)/(\wb z-\wh z), \\
\label{A-2-CAC-1 a}
 \wb{\wt \xi}=(y/x-\alpha_{3})\frac{\wt \xi-\wb \xi}{\wt z-\wb z}
+\frac{\alpha_{3}y-\eta}{x}-\alpha_{2}-\frac{G_{4}(-p,-a)\wt x-G_{4}(-r,-a)\wb x}{x(\wt z-\wb z)}, \\
\label{A-2-CAC-1 b}
 \wh{\wb \xi}=(y/x-\alpha_{3})\frac{\wb \xi-\wh \xi}{\wb z-\wh z}
+\frac{\alpha_{3}y-\eta}{x}-\alpha_{2}-\frac{G_{4}(-r,-a)\wb x-G_{4}(-q,-a)\wh x}{x(\wb z-\wh z)},\\
\label{A-2-CAC c}
 \wb{\wt{\eta}}= \Big(\wt{y}\wb{z}-\wb{y}\wt{z}-\frac{\xi}{z}(\wt{y}-\wb{y})\Big)/(\wt{z}-\wb{z}), \quad
\wb{\wh{\eta}}= \Big(\wh{y}\wb{z}-\wb{y}\wh{z}-\frac{\xi}{z}(\wh{y}-\wb{y})\Big)/(\wh{z}-\wb{z}).
\end{gather}
\end{subequations}
Furthermore, from the three right-hand side equations in \eqref{mat-F-abc} and using the data \eqref{A-2-CAC}, we
have three different ways to calculate the values $\wb{\wh{\wt{x}}},~\wb{\wh{\wt{y}}}$,~
$\wb{\wh{\wt{z}}}$ and $\wb{\wh{\wt{\eta}}}$ uniquely, which are
\begin{align*}
& \wb{\wh{\wt{x}}}= \big[(\wt{x}-\wh{x})\wb{z}+(\wh{x}-\wb{x})\wt{z}+(\wb{x}-\wt{x})\wh{z}\big]/
 \big[\big(\wt{\xi}-\wh{\xi}\big)\wb{z}+\big(\wh{\xi}-\wb{\xi}\big)\wt{z}+\big(\wb{\xi}-\wt{\xi}\big)\wh{z}\big], \\
& \wb{\wh{\wt{y}}}= \big[(\wh{x}-\wt{x})\wb{\xi}+(\wt{x}-\wb{x})\wh{\xi}+(\wb{x}-\wh{x})\wt{\xi}\big]/
 \big[(\wt{z}-\wh{z})\wb{\xi}+(\wh{z}-\wb{z})\wt{\xi}+(\wb{z}-\wt{z})\wh{\xi}\big], \\
& \wb{\wh{\wt{z}}}=\big[(\alpha_3x-y)\big((\wh{z}-\wb{z})\wt{\xi}+(\wb{z}-\wt{z})\wh{\xi}
+(\wt{z}-\wh{z})\wb{\xi}\big) \nn\\
&\hphantom{\wb{\wh{\wt{z}}}=}{}
+\wt{x}(\wh{z}-\wb{z})G_4(-p,-a) +\wh{x}(\wb{z}-\wt{z})G_4(-q,-a)+\wb{x}(\wt{z}-\wh{z})G_4(-r,-a)\big]\nn \\
& \qquad\qquad \qquad /\big[x\big((\wh{z}-\wt{z})\wb{\xi}
+(\wt{z}-\wb{z})\wh{\xi}+(\wb{z}-\wh{z})\wt{\xi}\big)\big], \\
&\wb{\wh{\wt{\eta}}}=\frac{ -\wb{\xi} \Big(\frac{\wt{x}\wb{z}-\wb{x}\wt{z}}{\wt{z}-\wb{z}}
-\frac{\wh{x}\wb{z}-\wb{x}\wh{z}}{\wh{z}-\wb{z}}\Big)
+\wb{z} \Big(\frac{(\wb{\xi}-\wh{\xi})(\wt{x}\wb{z}-\wb{x}\wt{z})}{(\wt{z}-\wb{z})(\wb{z}-\wh{z})}
-\frac{(\wb{\xi}-\wt{\xi})(\wh{x}\wb{z}-\wb{x}\wh{z})}{(\wh{z}-\wb{z})(\wb{z}-\wt{z})}\Big)}
{\wb{z} \Big(\frac{\wb{\xi}-\wt{\xi}}{\wb{z}-\wt{z}}-\frac{\wb{\xi}-\wh{\xi}}{\wb{z}-\wh{z}}\Big)}.
\end{align*}
For the variable $\xi$, the formula for $\wb{\wh{\wt{\xi}}}$ is too long to be listed here.
It turns out that to guarantee the relation
$\wb{\wh{\wt{\xi}}}=\wh{\wb{\wt{\xi}}}=\wt{\wb{\wh{\xi}}}$
holds one needs
\begin{equation*}
 \wt{\eta}\wh{x}-\wh{\eta}\wt{x}=y(\wt{x}-\wh{x}),\qquad
 \wb{\eta}\wh{x}-\wh{\eta}\wb{x}=y(\wb{x}-\wh{x}),\qquad
 \wt{\eta}\wb{x}-\wb{\eta}\wt{x}=y(\wt{x}-\wb{x})
\end{equation*}
hold. This is nothing but the second equation \eqref{eq:A2-ex_b} in \eqref{eq:A2-111}
and its $(\wb{\phantom{a}}, \wh{~~})$ and $(\wb{\phantom{a}}, \wt{~~})$ versions.
Thus we can conclude two points.
One is that although \eqref{eq:A2-ex_b}
is not needed for determining the up-right evolution of \eqref{eq:A2-111},
it is necessary for the whole system \eqref{eq:A2-111} to be MDC.
In other words, the GD-4 (A-2) equation \eqref{eq:A2-111} as a whole system satisfies the MDC property.
The other point is, if, without \eqref{eq:A2-ex_b} the system \eqref{eq:A2-111}
is integrable in the sense of being MDC, then \eqref{eq:A2-ex_b} is a consequence of the other five equations
in \eqref{eq:A2-111}.

For the alternative GD-4 (A-2) equation \eqref{eq:A2-222},
the GD-4 (B-2) equation \eqref{eq:B222},
the four-component GD-4 (A-2) equation \eqref{eq:4comp-A-21}
and its alternative form \eqref{eq:4comp-A-22},
their MDC property can be checked in a similar way and has similar results.
The check is straightforward and we present formulas of those triple shifts of variables in Appendix \ref{App-3}.
What we want to emphasize is,
for $\eta$ in the alternative GD-4 (A-2) equation \eqref{eq:A2-222},
$\xi$ in the GD-4 (B-2) equation \eqref{eq:B222},
$\xi$ in the four-component GD-4 (A-2) equation \eqref{eq:4comp-A-21},
and $\varpi$ in \eqref{eq:4comp-A-22},
their triple shifts are uniquely determined respective require
\eqref{eq:A2-ex-N4-1 bb}, \eqref{eq:BSQ-ex_b}, \eqref{eq:4comp-A-2 bb},
\eqref{eq:4comp-3.41b} and their $(\wb{\phantom{a}}, \wh{~~})$ and $(\wb{\phantom{a}}, \wt{~~})$
versions hold.
These facts provide supports for Remark \ref{R-4-2}.
On the other hand, if we agree that these systems are MDC, then
\eqref{eq:A2-ex-N4-1 bb}, \eqref{eq:BSQ-ex_bb}, \eqref{eq:4comp-A-2 bb} and
\eqref{eq:4comp-3.41b}
will be the consequence of the rest equations in their own system, respectively.

\subsection{Lax representations}\label{sec-5-3}

Finding a Lax pair for a given nonlinear equation, whether continuous or discrete, is generally
a~difficult task.
The DL scheme provides a direct
procedure to construct Lax pairs of the objective equations based on the relation~\eqref{eq:Nukrels}. One can select specific components of these
vectors, or combinations thereof, to constitute the basic vector functions
in terms of which we obtain the relevant linear problems. This has been illustrated
in the lattice KdV and DBSQ case (e.g.,~\cite{Walker}).
Besides, the fundamental characterization
of integrable partial difference equations as being MDC is intimately related to the
existence of a~Lax pair~\cite{ABS-2003,Bri,Nij-LP}. Always the information gained from
the process of verifying MDC property is also crucial to the
computation of the corresponding Lax pair.

For the quadrilateral ABS equations \cite{ABS-2003} and 3-point discrete Burgers equation \cite{CZZ-2021,Z-PEDAM-2022},
both of which are CAC,
they can be obtained as compatibilities of their Lax pairs constructed from the MDC property
(see~\cite{Bri,CZZ-2021,Z-PEDAM-2022}).
However, for a DBSQ type equation, e.g.,~(B-2) or ((A-2), (C-3)),
its Lax pairs, no matter constructed from the DL scheme or by means of the MDC property,
are incomplete in the sense that the original DBSQ equation can not be fully recovered from the compatibility
of its Lax pair (see \cite{HZ-2021,ZZN}).
Such an incompleteness of the Lax pair also happens to the GD-4 type equations.
In the following, we only take the GD-4 (A-2) equation \eqref{eq:A2-111} as an example and present its Lax pairs
derived from the DL scheme and the MDC property, respectively.
One can see that, the equations arising from the compatibility of the Lax pairs can not yield
all the equations in the GD-4 (A-2).

To determine the Lax pairs of the lattice equations using the DL approach, we should use
the relation~\eqref{eq:ukrels}. For the sake of the construction of the Lax pairs,
we introduce $(\mathbf{v})_0$ standing for the $0$-th component of the infinite-component column vector $\mathbf{v}$.
Besides, we appoint that $A^{T}$ stands for the transpose of matrix~$A$.

We start by looking at the Lax pair of \eqref{eq:A2-1}, which is the DL counterpart of the
GD-4 (A-2) equation~\eqref{eq:A2-111}.
We define an eigenvector of the form
\begin{align}\label{5.3}
\bphi=\big(\big((a+\Ld)^{-1}\bu_k\big)_0 ,(\bu_{k})_0,(\Ld\bu_{k})_0,\big(\Ld^2\bu_{k}\big)_0\big)^{\rm T}.
\end{align}
Based on \eqref{eq:ukrels}, one can derive the following linear relation
\begin{subequations}\label{eq:A2-Lax-f}
\begin{align}\label{eq:A2-Lax}
\wt{\bphi}=\bL\bphi,\qquad \wh{\bphi}=\bM\bphi,
\end{align}
in which
\begin{align}
\label{eq:A2-L}
\bL=\left(\begin{matrix}
p-a & \wt{v}_a & 0 & 0 \\
0 & p-\wt{u}_0 & 1 & 0 \\
0 & -\wt{u}_{1,0} & p & 1 \\
G_4(k,-a)/v_a & *_{42} &
(\alpha_3s_a-r_a)/v_a-\alpha_2 &
A_1(-p)+s_a/v_a
\end{matrix}\right),
\end{align}
where
\begin{gather}
*_{42}=p_a\wt{v}_a/v_a-\big[A_2(-p)+(A_1(-p)s_a+r_a)/v_a\big]
(p-\wt{u}_0)-\wt{u}_{1,0}(A_1(-p)+s_a/v_a),
\end{gather}
\end{subequations}
and $\bM$ is obtained from \eqref{eq:A2-L} by replacing $p$ by $q$
and $\wt{\phantom{a}}$ by $\wh{\phantom{a}}$.
The compatibility of \eqref{eq:A2-Lax}, i.e., $\wh{\bL} \bM =\wt{\bM} \bL$,
leads to equations \eqref{eq:bA2-b}, \eqref{eq:bA2-c},
\eqref{eq:bA2-d}, \eqref{eq:svurels_b} and
\begin{gather*}
p-q+\wh{s}_a/\wh{v}_a-\wt{s}_a/\wt{v}_a=v_a\big((p-a)/\wt{v}_a-(q-a)/\wh{v}_a\big).
\end{gather*}
The latter two equations follows readily from \eqref{eq:bA2-a}.
The incompleteness of the Lax pair is demonstrated by the missing of \eqref{eq:bA2-a}.

It is well known that the MDC property can provide Lax pairs automatically.
All the deformed lattice equations listed above admit the MDC property.
As an example, in this part
we discuss the Lax integrability of the GD-4 (A-2) equation \eqref{eq:A2-111} thanks to its MDC property.
As the discussion of the MDC property,
here we extend the aforesaid lattice equations into a third dimension by introducing
a new variable $l$ associated with a new shift $\wb{\phantom{a}}$
and a new complex parameter~$r$.
For the lattice equation~\eqref{mat-F} which is MDC,
the essential procedure to construct its Lax pair is to
introduce fractional expressions (e.g., $f/F,~g/G$, etc.) for the various components
in the third dimension in the B\"acklund transformation \eqref{mat-F-BT}
in order to linearize the numerators and denominators of the expressions for the components of
$\wt{\wb{\Tht}}$ and $\wh{\wb{\Tht}}$ in terms of~$f$,~$F$, $g$, $G$, etc. This manipulation
generates the Lax matrices.

For the GD-4 (A-2) equation \eqref{eq:A2-111},
we consider its B\"acklund transformation \eqref{A-2-CAC}.
Introducing
\begin{gather*}
\wb{x}=\frac{\varphi_1}{\varphi_0},\qquad \wb{z}=\frac{\varphi_2}{\varphi_0}, \qquad
\wb{y}=\frac{\varphi_3}{\varphi_0},\qquad \wb{\xi}=\frac{\varphi_4}{\varphi_0},\qquad
\wb{\eta}=\frac{\varphi_5}{\varphi_0},
\end{gather*}
and defining
\[\bphi_1=(\varphi_0, \varphi_1, \varphi_2, \varphi_3, \varphi_4, \varphi_5)^{\rm T},\]
from \eqref{A-2-CAC} we have
\begin{gather*}
\wt{\bphi}_1=\bL_1\bphi_1,
\qquad \wh{\bphi}_1=\bM_1 \bphi_1,
\end{gather*}
where
\begin{gather*}
\bL_1 =
\left(\begin{matrix}
\wt{z} & \hphantom{-}0 & -1 & 0 & \hphantom{-}0 & 0\\
\wt{x} & -1 & \hphantom{-}0 & 0 & \hphantom{-}0 & 0\\
\wt{\xi} & \hphantom{-}0 & \hphantom{-}0 & 0 & -1 & 0 \\
0 & -\wt{z} & \hphantom{-}\wt{x} & 0 & \hphantom{-}0 & 0 \\
*_{51} & G_4(-r,-a)/x & (\eta-\alpha_3y)/x+\alpha_2
& 0 & \alpha_3-y/x & 0 \\
-(\xi/z)\wt{y} & \hphantom{-}0 & \hphantom{-}\wt{y} & \xi/z-\wt{z} & \hphantom{-}0 & 0
\end{matrix}\right),
\end{gather*}
where $*_{51}=\wt{z}\big((\alpha_3y-\eta)/x-\alpha_2\big)+(y/x-\alpha_3)\wt{\xi}-G_4(-p,-a)\wt{x}/x$ and
the matrix $\bM_1$ is the hat-$q$ version of $\bL_1$.
The compatibility gives the relations:
\begin{gather}\label{eq:C3compat}
\wh{\wt x}=\frac{\wt x-\wh x}{\wt z-\wh z}, \qquad \wh{\wt{z}}=\frac{\wh{\xi}-\wt{\xi}}{\wh{z}-\wt{z}},
 \qquad x=\frac{\wh x \wt y-\wt x\wh y}{\wt x-\wh x},
 \qquad \wh{\wt y}=\frac{\wh x \wt z-\wt x \wh z}{\wh z-\wt z},
 \qquad y=\frac{\wt x \wh \eta-\wh x \wt \eta}{\wh x-\wt x},
\end{gather}
together with \eqref{eq:A2-ex_c}. The second equation in \eqref{eq:C3compat} is \eqref{eq:A2-ex_b},
while the rest four equations in~\eqref{eq:C3compat}
are consequences of the equations \eqref{eq:A2-ex_a} and \eqref{eq:A2-ex_b} in the GD-4 (A-2) equation.
Similar to the DBSQ case \cite{ZZN}, here we can
remove the last columns and rows in $\bL_{1}$ and $\bM_{1}$, and examine the remains:
\begin{gather}\label{eq:A2-LP-4O}
\wt{\bphi}_{2}=\bL_{2}\bphi_{2}, \qquad \wh{\bphi}_{2}=\bM_{2}\bphi_{2},
\end{gather}
in which $\bphi_{2}=(\varphi_0, \varphi_1, \varphi_2, \varphi_3, \varphi_4)^{\rm T}$ and
\begin{gather*}
\bL_{2}=\left(
\begin{matrix}
\wt{z} & \hphantom{-}0 & -1 & 0 & \hphantom{-}0 \\
\wt{x} & -1 & \hphantom{-}0 & 0 & \hphantom{-}0 \\
\wt{\xi} & \hphantom{-}0 & \hphantom{-}0 & 0 & -1 \\
0 & -\wt{z} & \hphantom{-}\wt{x} & 0 & \hphantom{-}0 \\
*_{51} & G_4(-r,-a)/x & (\eta-\alpha_3y)/x+\alpha_2
& 0 & \alpha_3-y/x
\end{matrix}\right),
\end{gather*}
where the matrix $\bM_{2}$ is the hat-$q$ version of
$\bL_{2}$. The compatibility of \eqref{eq:A2-LP-4O} gives
\eqref{eq:C3compat} and~\eqref{eq:A2-ex_c} as well.

The above are Lax pairs of the GD-4 (A-2) equation.
Similarly, we can get Lax pairs for other GD-4 type lattice equations we have obtained.
However, all the Lax pairs are incomplete. In other words, not all the equations in the GD-4 type equations
can be obtained from the compatibility
of their Lax pairs.

\section{Conclusions and remarks}\label{sec-6}

In this paper, we have analyzed the construction of the extended lattice GD type equations with $N=4$ by DL approach.
We first reviewed the DL scheme for the extended generic lattice GD type equations, which appeared in~\cite{ZZN}.
Then we restricted ourselves to the case of $N=4$ and showed the derivation of the lattice GD-4 type equations, including
the lattice GD-4 (A-2), (B-2) and (C-3) equations.
In the DL scheme, these equations appear as closed forms, each of which
includes five (independent) variables.
We introduced point transformations to convert them into Hietarinta's version,
which are the GD-4 counterparts of the DBSQ type equations (see Section~\ref{sec-4-2}).
We also obtained reduced GD-4 (A-2) equations with four variables.
In addition, we investigated the evolutions and MDC property of these equations.
As examples, for the GD-4 (A-2) equation we gave its Lax pairs
constructed from the DL scheme as well as the MDC property.
One can also see that some lower equations
are embedded into the GD-4 type equations due to the coefficients $\alpha_3$ and~$\alpha_2$
in $G_4(\omega,k)=0$. For example, in the GD-4 (B-2) equation~\eqref{eq:B222},
equation \eqref{eq:BSQ-ex_a} and the coefficients of $\alpha_3$ and $\alpha_2$ in~\eqref{eq:BSQ-ex_c}
compose the DBSQ (B-2) equation (see~\eqref{eq:BSQ-ex-N=3}),
while the coefficient of $\alpha_2$ in~\eqref{eq:BSQ-ex_c}
gives rise to the lattice potential KdV equation (i.e., H1 equation in the ABS list)
$\big(x-\widehat{\widetilde{x}}\big)(\wh x- \wt x)=p^2-q^2$.

We finish the paper with the following remarks.
First of all, compared with the DBSQ case ($N=3$, see \cite{H,HZ-2021,ZZN} and also Appendix~\ref{App-1}) which are all MDC,
the GD-4 type lattice equations have something special.
One needs to introduce more variables to obtain quadrilateral closed forms in the DL scheme.
All the obtained lattice GD-4 equations allow up-left as well as down-right evolutions
with staircase initial values,
but not all of them allow up-right evolution.
For the GD-4 (B-2) equation, we introduced the sixth equation \eqref{eq:u11-can_cd}
(or~\eqref{eq:B2-ex_eta-th} or~\eqref{eta12} in Hietarinta's form
so that the equation set gets an up-right evolution.
The enlarged system~\eqref{eq:B222} therefore contains six equations but with five variables.
However, all six equations are necessary to guarantee the system to be MDC.
Among the six equations in the system \eqref{eq:B222},
equation~\eqref{eq:BSQ-ex_bb} is defined on the triangle \smash[b]{$\big(V,\wt{V},\wh{V}\big)$}
(not defined on the edge, cf.~\eqref{eq:BSQ-ex_a}),
which has zero contribution in the up-right evolution, but
is necessary for the relation \smash[b]{$\wb{\wh{\wt{\xi}}}=\wh{\wb{\wt{\xi}}}=\wt{\wb{\wh{\xi}}}$}
to be held
(see~\eqref{C.2} in Appendix~\ref{App-3}).
On the other side, this fact also means that, if we admit that the GD-4 (B-2) system \eqref{eq:B222}
excluding \eqref{eq:BSQ-ex_bb} is integrable, which indicates the system is MDC,
then \eqref{eq:BSQ-ex_bb} will be a consequence of the other five equations in system \eqref{eq:B222}.
Similar feature holds for the GD-4 (A-2) equation, its alternative form and their four-component versions.
Note that the MDC property has recently been used to construct B\"acklund transformation and
find one-soliton solution to the GD-4 (B-2) equation, see~\cite{TZ-2023}.
Secondly, it is not clear how such GD-4 lattice equations are related to the discrete KP-type equations.
In~\cite{Atk}, the lattice modified and lattice Schwarzian GD hierarchies were obtained
as reductions of the discrete KP system (see \cite[equations~(C.7) and~(C.12)]{Atk},
and see also~\cite{ALN}),
and those equations are presented in quadrilateral form in terms of multi-components.
For the GD-4 case, they are 3-component quad systems.
It remains open that how they are related to our GD-4 (A-2) and (C-3) equations.
What's more, compared with the DBSQ equations (cf.~\cite{HZ-2021}),
there are still many topics to be considered in order to understand the GD-4 type lattice equations.
For example,
first, the one-component forms of these equations need to be determined.
They are supposed to live on a~$4\times 4$ (16-point) stencil, considering the
$(\,\wt{~},p) \to (\,\wh{~},q)$ symmetry.
However, it is not easy to eliminate auxiliary variables to get an equation for a single component,
(compared with the DBSQ case in \cite{HZ-2021}). We will leave this as a topic in the next step.
Besides, one can choose specific measures~$d\ld_j(k)$ and integration contour $\Gamma_j$ in \eqref{eq:bC}
to get soliton solutions, which can be operated along the lines of~\cite{ZZN}.
In addition, we have not touched the $\tau$-function, which in the case
of $N=3$ satisfies a trilinear equation (see~\cite{ZZN} or~\cite{HZ-2021}).
It is undoubtedly that figuring out the $\tau$-function of the lattice GD-4 case and its
bilinear equations (cf.~\cite{HJN-book,HZ-2009,HZ-JMP-2010,HZ-SIGMA-2011,ZZ-JNMP}
for the ABS and DBSQ equations) is meaningful and worth considering.

\appendix

\section{The DBSQ-type equations}\label{App-1}

The DBSQ-type equations, including (A-2), (B-2),
(C-3) and (C-4) equations \cite{H}, which are related to the third-order polynomial \cite{ZZN}
\begin{gather*}
G_3(\oa,k):=\oa^3-k^3+\alpha_2\big(\oa^2-k^2\big)+\alpha_1(\oa-k),
\end{gather*}
are listed as follows.

{\bf (A-2) equation}: This equation has two forms. One is
\begin{gather*}
 \wt{y}=z\wt{x}-x, \qquad \wh{y}=z\wh{x}-x, \qquad
 y =x\wh{\wt{z}}-b_0x- \big(G_3(-p,-a)\,\wt{x}-G_3(-q,-a)\,\wh{x}\big)/(\wh{z}-\wt{z}),
\end{gather*}
and the other is
\begin{gather*}
 y=\wt{z}x-\wt{x}, \qquad y=\wh{z}x-\wh{x}, \qquad
\wh{\wt{y}}=\wh{\wt{x}}z-b'_0\wh{\wt{x}}- \big(G_3(-p,-b)\wh{x}-G_3(-q,-b)\wt{x}\big)/(\wh{z}-\wt{z}).
\end{gather*}
Here $b_0$ and $b'_0$ are constants relating to parameter $\alpha_2$.
Both of these two equations are associated with each other
by reversal symmetry
\[p\rightarrow -p, \qquad q\rightarrow -q, \qquad n\rightarrow -n, \qquad m\rightarrow -m, \qquad
\alpha_2\rightarrow -\alpha_2, \qquad b \rightarrow -a. \]

{\bf (B-2) equation}: This equation reads
\begin{subequations}\label{eq:BSQ-ex-N=3}
\begin{gather}
 \wt{z}=x\wt{x}-y, \quad \wh{z}=x\wh{x}-y, \\
 z=x\wh{\wt{x}}-\wh{\wt{y}}-\alpha_2\big(\wh{\wt{x}}-x\big)-\alpha_1-G_3(-p,-q)/(\wh{x}-\wt{x}).
\end{gather}
\end{subequations}
The parameter $\alpha_1$ can be removed by transformation.

{\bf (C-3) equation}: One equation of this type is
\begin{subequations}
\label{eq:xyz-MSBSQ-app}
\begin{gather}
 x-\wt{x}=\wt{y}z, \qquad x-\wh{x}=\wh{y}z, \\
 y \wh{\wt{z}}=z \big(G_3(-p,-b)\wh{z}\wt{y}-G_3(-q,-b)\wt{z}\wh{y}\big)/(\wh{z}-\wt{z})
+G_3(-a,-b)\wh{\wt{x}},
\end{gather}
\end{subequations}
and the other is
\begin{subequations}\label{eq:xyz-MSBSQ-1-app}
\begin{gather}
\label{eq:xyz-MSBSQ-1a-app}
 x-\wt{x}=\wt{y}z, \qquad x-\wh{x}=\wh{y}z, \\
\label{eq:xyz-MSBSQ-1b-app}
 y \wh{\wt{z}}=z \big(G_3(-p,-a)\wh{z}\wt{y}-G_3(-q,-a)\wt{z}\wh{y}\big)/(\wh{z}-\wt{z})+G_3(-a,-b)x.
\end{gather}
\end{subequations}
The equations \eqref{eq:xyz-MSBSQ-app} and \eqref{eq:xyz-MSBSQ-1-app} are
related by reversal symmetry
\[n\rightarrow -n, \qquad m\rightarrow -m \qquad y\rightarrow z, \qquad z\rightarrow -y, \qquad a \leftrightarrow b.\]

{\bf (C-4) equation}: This equation arises from the combination of equations
\eqref{eq:xyz-MSBSQ-app} and \eqref{eq:xyz-MSBSQ-1-app}
\begin{subequations}\label{eq:C4-N=3}
\begin{gather}
\label{eq:C4-N=3-a}
 x-\wt{x}=\wt{y}z, \qquad x-\wh{x}= \wh{y}z, \\
\label{eq:C4-N=3-b}
 y \wh{\wt z}=z\,\big(P'_{a,b}\,\wh{z}\wt{y}-Q'_{a,b}\,\wt{z}\wh{y}\big)/(\wh{z}-\wt{z})+
x \wh{\wt x}-G'^2_{a,b},
\end{gather}
\end{subequations}
where
\begin{gather*}
P'_{a,b}= \big(G_3(-p,-b)+G_3(-p,-a)\big)/2,~Q'_{a,b}= \big(G_3(-q,-b)+G_3(-q,-a)\big)/2,
~G'_{a,b}\\
\hphantom{P'_{a,b}}{}=G_3(-a,-b)/2.
\end{gather*}

\section{Derivation of some equations in Section \ref{sec-3-1}}\label{App-2}

\subsection{Equation (\ref{eq:ssrels_b})}\label{App-2-1}

For the both sides of equation \eqref{eq:Urels_b},
we multiply from the left by $\tbme(a+\Ld)^{-1}$ and from the right by $\big({-}b+\!\tLd\big)^{-1}\bme$.
For the left-hand side, we have
\begin{equation}\label{B-1}
\tbme (a+\Ld)^{-1}\bU\big(A_3(-p)+A_2(-p)\tLd +A_1(-p)\tLd^{2} +\tLd^{3}\big)\big({-}b+\!\tLd\big)^{-1}\bme.
\end{equation}
Calculating it term by term and express the results in terms of the variables introduced in \eqref{eq:objs},
we have
\begin{gather*}
A_3(-p)\tbme (a+\Ld)^{-1}\bU\big({-}b+\!\tLd\big)^{-1}\bme = A_3(-p)s_{a,b},\\
 A_2(-p)\tbme (a+\Ld)^{-1}\bU\tLd\big({-}b+\!\tLd\big)^{-1}\bme \\
\qquad {}= A_2(-p)\tbme (a+\Ld)^{-1}\bU\big(\tLd-b+b\big)\big({-}b+\!\tLd\big)^{-1}\bme \\
\qquad {}= A_2(-p)\big[\tbme (a+\Ld)^{-1}\bU \bme+b \tbme (a+\Ld)^{-1}U \big({-}b+\!\tLd\big)^{-1}\bme\big] \\
\qquad {}= -A_2(-p)(v_{a}-1)+ bA_2(-p)s_{a,b},
\\
A_1(-p)\tbme (a+\Ld)^{-1}\bU\tLd^{2}\big({-}b+\!\tLd\big)^{-1}\bme \\
\qquad {}= A_1(-p)\tbme (a+\Ld)^{-1}\bU\big(\tLd^{2}-b^2+b^2\big)\big({-}b+\!\tLd\big)^{-1}\bme \\
\qquad {}= A_1(-p)\big[\tbme (a+\Ld)^{-1}\bU\big(\tLd+b\big)\bme
+b^2\tbme (a+\Ld)^{-1}\bU \big({-}b+\!\tLd\big)^{-1}\bme\big] \\
\qquad {}= A_1(-p)[(a-s_{a})-b(v_{a}-1)]+b^{2}A_1(-p) s_{a,b},
\end{gather*}
and
\begin{gather*}
 \tbme (a+\Ld)^{-1}\bU\tLd^{3}\big({-}b+\!\tLd\big)^{-1}\bme \\
\qquad {}= \tbme (a+\Ld)^{-1}\bU\big(\tLd^{3}-b^3+b^3\big)\big({-}b+\!\tLd\big)^{-1}\bme \\
\qquad {}= \tbme (a+\Ld)^{-1}\bU\big(\tLd^{2}+ b \tLd+b^2\big)\bme
+b^3\, \tbme (a+\Ld)^{-1}\bU\big({-}b+\!\tLd\big)^{-1}\bme \\
\qquad {}= a^{2}-r_{a}+b(a-s_{a})-b^{2}(v_{a}-1)+b^{3}s_{a,b}.
\end{gather*}
Thus, equation \eqref{B-1} yields
\begin{gather*}
\big[A_2(-p)+(a+b)A_1(-p)+a^2+ab+b^2\big]+p_bs_{a,b}\\
\qquad{}-A_2(-p)v_{a}- A_1(-p)(s_{a}+b v_{a})-r_{a}-b s_{a}-b^{2}v_{a},
\end{gather*}
where $p_b$ is defined as in \eqref{eq:papb}.

For the right-hand side, we can calculate
\[
\tbme (a+\Ld)^{-1}\big(A_3(-p)-A_2(-p)\Ld +A_1(-p)\Ld^{2} -\Ld^{3}\big)\wt\bU\big({-}b+\!\tLd\big)^{-1}\bme
\]
in a similar way, which gives rise to
\begin{gather*}
-\big[A_2(-p)+(a+b)A_1(-p)+a^2+ab+b^2\big]+p_a \wt s_{a,b} \\
\qquad{}+A_2(-p)\wt w_{b}- A_1(-p)\big(\wt t_{b}-a \wt w_{b}\big)+\wt z_{b}-a \wt t_{b} +a^{2}\wt w_{b},
\end{gather*}
where $p_a$ is defined as in \eqref{eq:papb}.
For those terms in \eqref{eq:Urels_b} where $\bO$ is involved, for example,
$-\bU\tLd \bO\Ld\wt{\bU}$, we make use of $\bO= \bme \tbme$ and have
\begin{gather*}
 - \tbme (a+\Ld)^{-1}\bU\tLd \bO\Ld\wt{\bU} \big({-}b+\!\tLd\big)^{-1}\bme \\
\qquad{}= - \tbme (a+\Ld)^{-1}\bU \tLd \bme \cdot \tbme \wt{\bU} \big({-}b+\!\tLd\big)^{-1}\bme
= -(s_{a}-a)(1-\wt{w}_{b}).
\end{gather*}
 We can combine all these results together. Finally, we arrive at equation \eqref{eq:ssrels_b}.

\subsection{Equation (\ref{eq:fgrels_a})}\label{App-2-2}

For the both sides of equation~\eqref{eq:Urels_a},
multiplying from the left by $\tbme (a+\Ld)^{-1}$ and from the right by $\tLd^{2}\bme$,
we have
\begin{gather}
\tbme (a+\Ld)^{-1} \wt{\bU} \big(p- \! \tLd\big)\, \tLd^{2}\bme\nonumber\\
\qquad{}
=\tbme (a+\Ld)^{-1} (p+\Ld) \bU\,\tLd^2 \bme
-\tbme (a+\Ld)^{-1} \wt{\bU} \bO \bU\,\tLd^2 \bme.\label{B-2}
\end{gather}
The left-hand side of the above equation gives rise to
\begin{gather*}
\eqref{B-2}|_{\mathrm{l.h.s}} = p \,\tbme (a+\Ld)^{-1} \wt{\bU}\, \tLd^{2}\bme
-\tbme (a+\Ld)^{-1} \wt{\bU}\,\tLd^{3}\bme\\
\hphantom{\eqref{B-2}|_{\mathrm{l.h.s}}}{}
=p a^2-p\wt r_a+\wt f_a-a^3,
\end{gather*}
and the right-hand side yields
\begin{gather*}
\eqref{B-2}|_{\mathrm{r.h.s}}
 = \tbme\,(a+\Ld)^{-1} (p-a+a+\Ld) \bU\,\tLd^2 \bme-
\tbme (a+\Ld)^{-1} \wt{\bU} \bme \cdot \tbme \bU\,\tLd^2 \bme\\
\hphantom{\eqref{B-2}|_{\mathrm{r.h.s}}}{} = (p-a)\tbme\,(a+\Ld)^{-1} \bU\,\tLd^2 \bme+\tbme \bU\,\tLd^2 \bme
+(\wt{v}_{a}-1)u_{0,2}\\
\hphantom{\eqref{B-2}|_{\mathrm{r.h.s}}}{} = (p-a) \big(a^2- r_a\big)+u_{0,2}+(\wt{v}_{a}-1) u_{0,2}\\
\hphantom{\eqref{B-2}|_{\mathrm{r.h.s}}}{}=(p-a) \big(a^2- r_a\big)+\wt{v}_{a} u_{0,2}.
\end{gather*}
Equation \eqref{eq:fgrels_a} follows from the combination of the above results.

\subsection{Equation (\ref{eq:gfrels_a})}\label{App-2-3}

To obtain equation \eqref{eq:gfrels_a}, we multiply \eqref{eq:Urels_b} from the left
by $\tbme (a+\Ld)^{-1}$ and from the right by $\tLd^{2}\bme$.
The left-hand side yields
\begin{gather*}
 \tbme (a+\Ld)^{-1}\bU\big(A_3(-p)+A_2(-p)\tLd +A_1(-p)\tLd^{2} +\tLd^{3}\big) \bme \\
\qquad{}= A_3(-p) (1-v_a) +A_2(-p) (a-s_a) +A_1(-p) \big(a^2-r_a\big) +\big(a^3-f_a\big)\\
\qquad{} = p_a -A_3(-a) v_a- A_2(-p) s_a -A_1(-p) r_a -f_a,
\end{gather*}
where $p_a$ is defined as in \eqref{eq:papb}.
The right-hand side yields
\begin{gather*}
 \tbme (a+\Ld)^{-1}\bigl [ \big(A_3(-p)-A_2(-p)\Ld+A_1(-p)\Ld^2-\Ld^3\big)\wt{\bU} \nn \\
\qquad\quad{} +\bU \big[ A_2(-p)\bO
 -A_1(-p)\big(\bO \Ld-\tLd \bO\big)+\big(\bO \Ld^2-\tLd \bO \Ld+\tLd^2 \bO\big)\big]\wt{\bU}\bigr]\bme \\
\qquad{} = A_3(-p) \tbme (a+\Ld)^{-1}\wt{\bU}\bme
 -A_2(-p)\tbme (a+\Ld)^{-1}(\Ld+a-a)\wt{\bU}\bme \\
\qquad\quad{} +A_1(-p)\tbme (a+\Ld)^{-1}\big(\Ld^2-a^2 +a^2\big)\wt{\bU}\bme
 + \tbme (a+\Ld)^{-1}\big(\Ld^3+a^3 -a^3\big)\wt{\bU}\bme \\
\qquad\quad{}+A_2(-p) \tbme (a+\Ld)^{-1} \bU \bme \cdot \tbme \wt{\bU}\bme
 -A_1(-p) \tbme (a+\Ld)^{-1} \bU \bme \cdot \tbme \Ld \wt{\bU}\bme \\
\qquad\quad{}+A_1(-p) \tbme (a+\Ld)^{-1} \bU \tLd \bme \cdot \tbme \wt{\bU}\bme
 + \tbme (a+\Ld)^{-1} \bU \bme \cdot \tbme \Ld^2 \wt{\bU}\bme \\
\qquad\quad{}- \tbme (a+\Ld)^{-1} \bU \tLd \bme \cdot \tbme \Ld \wt{\bU}\bme
 + \tbme (a+\Ld)^{-1} \bU \tLd^2 \bme \cdot \tbme \wt{\bU}\bme \\
\qquad{} = A_3(-p) (1-\wt v_a) -A_2(-p) \wt u_0 +a A_2(-p) (1-\wt v_a) + A_1(-p) \wt u_{1,0}
\\
\qquad\quad{} - a A_1(-p) \wt u_{0} +a^2 A_1(-p) (1-\wt v_a)- \wt u_{2,0} +a \wt u_{1,0} - a^2 \wt u_{0} \\
\qquad\quad{}+a^3 (1-\wt v_a) + A_2(-p) (1-v_a) \wt u_{0}- A_1(-p) (1-v_a) \wt u_{1,0} \\
\qquad\quad{}+ A_1(-p) (a-s_a) \wt u_{0} + (1-v_a) \wt u_{2,0}-(a-s_a) \wt u_{1,0}+\big(a^2-r_a\big) \wt u_{0} \\
\qquad {}= p_a- p_a \wt v_a-v_a( A_2(-p) \wt u_{0}-A_1(-p) \wt u_{1,0}+\wt u_{2,0})
 - s_a(A_1(-p) \wt u_{0}-\wt u_{1,0}) -r_a \wt u_{0}.
\end{gather*}
Combining them together we get \eqref{eq:gfrels_a}.

\section{MDC property of the GD-4 (A-2) and (B-2) equations}\label{App-3}

We present procedure of checking MDC property of
the alternative GD-4 (A-2) equation \eqref{eq:A2-222}.
Since the checking is straightforward, for
the GD-4 (B-2) equation \eqref{eq:B222},
the four-component GD-4 (A-2) equation \eqref{eq:4comp-A-21}
and its alternative form \eqref{eq:4comp-A-22},
we only list out the triple shifts of the involved variables.

{\textbf{Alternative GD-4 (A-2) equation \eqref{eq:A2-222}}}:
We have
\begin{subequations}
\label{alt-A-2-CAC}
\begin{gather}
\label{alt-A-2-CAC a}
 \wb{\wt x}=(\wt{x}\wb{y}-\wb{x}\wt{y})/(\wb{x}-\wt{x}),
\qquad \wb{\wt y}=(\wt x \wb \eta-\wb x \wt \eta)/(\wb x-\wt x),
\qquad
\wb{\wt z}=(\wt y-\wb y)/(\wt x-\wb x), \\
\label{alt-A-2-CAC b}
 \wb{\wh x}=(\wh{x}\wb{y}-\wb{x}\wh{y})/(\wb{x}-\wh{x}),
\qquad \wb{\wh y}=(\wh x \wb \eta-\wb x \wh \eta)/(\wb x-\wh x), \qquad
\wb{\wh z}=(\wh y-\wb y)/(\wh x-\wb x), \\
\label{alt-A-2-CAC-1 a}
 \wb{\wt \xi}=\frac{\eta \wt x-y \wt y-(\eta +\wt y)\wb x+(y+\wt x)\wb y} {x(\wb x-\wt x)},
\quad \frac{\eta \wh x-y \wh y-(\eta +\wh y)\wb x+(y+\wh x)\wb y} {x(\wb x-\wh x)},\\
\label{alt-A-2-CAC-2 a}
 \wb{\wt \eta}=(\alpha_{3} z-\xi -\alpha_{2})\frac{\wt{x}\wb{y}-\wb{x}\wt{y}}{\wb{x}-\wt{x}}
+(z-\alpha_{3})\frac{\wt x \wb \eta-\wb x \wt \eta}{\wb x-\wt x}
-\frac{G4(-p,-b)\wb{x}-G4(-r,-b)\wt{x}}{\wb{z}-\wt{z}}, \\
\label{alt-A-2-CAC-2 b}
 \wb{\wh \eta}=(\alpha_{3} z-\xi -\alpha_{2})\frac{\wh{x}\wb{y}-\wb{x} \wh{y}}{\wb{x}-\wh{x}}
+(z-\alpha_{3})\frac{\wh x \wb \eta-\wb x \wh \eta}{\wb x-\wh x}
-\frac{G4(-q,-b)\wb{x}-G4(-r,-b)\wh{x}}{\wb{z}-\wh{z}}.
\end{gather}
\end{subequations}
Furthermore, from the three right-hand side equations in \eqref{mat-F-abc} and using the data \eqref{alt-A-2-CAC}
we have three different ways to calculate the values $\wb{\wh{\wt{x}}}$, $\wb{\wh{\wt{y}}}$,
$\wb{\wh{\wt{z}}}$ and $\wb{\wh{\wt{\xi}}}$ uniquely, which are
\begin{gather*}
 \wb{\wh{\wt{x}}}= \big[(\wh{x}\wb{\eta}-\wb{x}\wh{\eta})\wt{y}
+(\wb{x}\wt{\eta}-\wt{x}\wb{\eta})\wh{y}+(\wt{x}\wh{\eta}-\wh{x}\wt{\eta})\wb{y}\big]/
 \big[(\wh{x}-\wb{x})\wt{y}+(\wb{x}-\wt{x})\wh{y}+(\wt{x}-\wh{x})\wb{y}\big], \\
 \wb{\wh{\wt{z}}}= \big[(\wh{x}-\wt{x})\wb{\eta}+(\wt{x}-\wb{x})\wh{\eta}
+(\wb{x}-\wh{x})\wt{\eta}\big]/
 \big[(\wt{y}-\wh{y})\wb{x}+(\wh{y}-\wb{y})\wt{x}+(\wb{y}-\wt{y})\wh{x}\big],\\
 \wb{\wh{\wt{\xi}}}=
B_1 \times\left[\frac{(\wt{x}\wb{y}-\wb{x}\wt{y})\wb{\eta}}{\wb{x}-\wt{x}}
-\frac{(\wb{x}\wt{\eta}-\wt{x}\wb{\eta})\wb{y}}{\wt{x}-\wb{x}}
-\frac{\big(\wb{\eta}+\frac{\wb{x}\wt{\eta}-\wt{x}\wb{\eta}}
{\wt{x}-\wb{x}}\big)(\wh{x}\wb{y}-\wb{x}\wh{y})}{\wb{x}-\wh{x}}
+\frac{\big(\wb{y}+\frac{\wt{x}\wb{y}-\wb{x}\wt{y}}{\wb{x}-\wt{x}}\big)(\wb{x}\wh{\eta}-\wh{x}\wb{\eta})}
{\wh{x}-\wb{x}}\right],\\
 \wb{\wh{\wt{y}}}=\frac{B_2\!+B_3\!+x \big( (\wb{y}\!-\!\wt{y})y+(\wt{y}\wb{z}\!
-\!\wb{y}\wt{z})x\big)G_{4}(-q,-b)+x \big( (\wt{y}\!-\!\wh{y})y+(\wh{y}\wt{z}\!
-\!\wt{y}\wh{z})x\big)G_{4}(-r,-b)}{x \big((\wh{z}-\wt{z})\wb{y}+(\wt{z}-\wb{z})\wh{y}
+(\wb{z}-\wh{z})\wt{y}\big)},
\end{gather*}
where
\begin{gather*}
 B_1= \frac{1}{\big(\frac{\wt{x}\wb{y}-\wb{x}\wt{y}}{\wb{x}-\wt{x}}
+\frac{\wh{x}\wb{y}-\wb{x}\wh{y}}{\wb{x}-\wh{x}}\big)\wb{x}},\qquad
 B_2= x \big( (\wh{y}-\wb{y})y+(\wb{y}\wh{z}-\wh{y}\wb{z})x\big)G_{4}(-p,-b),\\
 B_3=(\alpha_{3}-z) \big[\wb{\eta} \big( (\wt{y}-\wh{y})y+(\wt{y}\wh{z}-\wh{y}\wt{z})x)\big)
+\wh{\eta} \big( (\wt{y}-\wb{y})y+(\wb{y}\wt{z}-\wt{y}\wb{z})x)\big) \\
\hphantom{B_3=}{}+\wt{\eta} \big( (\wb{y}-\wh{y})y+(\wh{y}\wb{z}-\wb{y}\wh{z})x)\big)\big].
\end{gather*}
For the variable $\eta$, the formula for $\wb{\wh{\wt{\eta}}}$ is somewhat long to be listed here.
One can check the coincide relation $\wb{\wh{\wt{\eta}}}=\wh{\wb{\wt{\eta}}}=\wt{\wb{\wh{\eta}}}$
by means of mathematical softwares, e.g., \textsc{Mathematica},
which holds only if \eqref{eq:A2-ex-N4-1 bb} and
its $(\wb{\phantom{a}}, \wh{~~})$ and $(\wb{\phantom{a}}, \wt{~~})$ versions hold,
i.e.,
\[ \wh{\xi}-\wt{\xi}=(\wh{z}-\wt{z})z,\qquad
\wb{\xi}-\wt{\xi}=(\wb{z}-\wt{z})z,\qquad
\wh{\xi}-\wb{\xi}=(\wh{z}-\wb{z})z. \]
In conclusion,
the alternative GD-4 (A-2) equation \eqref{eq:A2-222} as a system is MDC;
although \eqref{eq:A2-ex-N4-1 bb} is not needed for determining the up-right evolution of~\eqref{eq:A2-222},
it is necessary for the whole system \eqref{eq:A2-222} to be MDC.

{\textbf{GD-4 (B-2) equation \eqref{eq:B222}}}:
We have unique expressions for the values $\wb{\wh{\wt{x}}}$, $\wb{\wh{\wt{z}}}$,
$\wb{\wh{\wt{\eta}}}$ and $\wb{\wh{\wt{y}}}$, which are
\begin{gather*}
 \wb{\wh{\wt{x}}}=\big[(\wh{x}-\wt{x})\wb{\xi}+(\wt{x}-\wb{x})\wh{\xi}+(\wb{x}-\wh{x})\wt{\xi}\big]/
\big[(\wt{y}-\wh{y})\wb{x}+(\wh{y}-\wb{y})\wt{x}+(\wb{y}-\wt{y})\wh{x}\big], \\
 \wb{\wh{\wt{z}}}=\big[(\wh{y}-\wt{y})\wb{\xi}+(\wt{y}-\wb{y})\wh{\xi}+(\wb{y}-\wh{y})\wt{\xi}\big]/
[(\wt{y}-\wh{y})\wb{x}+(\wh{y}-\wb{y})\wt{x}+(\wb{y}-\wt{y})\wh{x}], \\
 \wb{\wh{\wt{\eta}}}=\frac{\big(\wh{x}\wb{\xi}-\wb{x}\wh{\xi}\big)\wt{y}+\big(\wb{x}\wt{\xi}
-\wt{x}\wb{\xi}\big)\wh{y}+\big(\wt{x}\wh{\xi}-\wh{x}\wt{\xi}\big)\wb{y}}
{(\wh{x}-\wb{x})\wt{y}+(\wb{x}-\wt{x})\wh{y}+(\wt{x}-\wh{x})\wb{y}}, \\
 \wb{\wh{\wt{y}}}=\frac{1}{\frac{\wt{y}-\wb{y}}{\wt{x}-\wb{x}}+\frac{\wb{y}-\wh{y}}{\wh{x}-\wb{x}}}
\bigg[\frac{G_{4}(-p,-r)+(x-\alpha_{3})\big(\wt{\xi}-\wb{\xi}\big)-(\wt{y}-\wb{y})(\alpha_{2}
-\alpha_{3}x+z)}{\wt{x}-\wb{x}} \nn \\
\hphantom{\wb{\wh{\wt{y}}}=}{}
+\frac{-G_{4}(-q,-r)+(-x+\alpha_{3})\big(\wh{\xi}-\wb{\xi}\big)+(\wh{y}-\wb{y})
(\alpha_{2}-\alpha_{3}x+z)}{\wh{x}-\wb{x}}\bigg].
\end{gather*}
The formula for $\wb{\wh{\wt{\xi}}}$ is too long to be listed here.
The coincident relation $\wb{\wh{\wt{\xi}}}=\wh{\wb{\wt{\xi}}}=\wt{\wb{\wh{\xi}}}$
holds only if~\eqref{eq:BSQ-ex_bb} and
its $(\wb{\phantom{a}}, \wh{~~})$ and $(\wb{\phantom{a}}, \wt{~~})$ versions hold,
i.e.,
\begin{equation}\label{C.2}
\wh{\eta}-\wt{\eta}=(\wh{x}-\wt{x})z,\qquad
\wb{\eta}-\wt{\eta}=(\wb{x}-\wt{x})z,\qquad
\wh{\eta}-\wb{\eta}=(\wh{x}-\wb{x})z.
\end{equation}
Thus,
the GD-4 (B-2) equation \eqref{eq:B222} as a system is MDC;
\eqref{eq:BSQ-ex_bb} is not needed for determining the up-right evolution of \eqref{eq:B222},
but it is necessary for the whole system \eqref{eq:B222} to be MDC.

{\textbf{Four-component GD-4 (A-2) equation \eqref{eq:4comp-A-21}}}:
$\wb{\wh{\wt{u}}}$, $\wb{\wh{\wt{z}}}$ and $\wb{\wh{\wt{v}}}$ are uniquely expressed as
\begin{gather*}
 \wb{\wh{\wt{u}}}=\big[\big(\wt{\xi}-\wh{\xi}\big)z\wb{u}+\wt{u}\big(-\wb{u}\wt{\xi}+\wh{u}\big(\wt{\xi}-\wh{\xi}\big)
+\wb{u}\wb{\xi}+\wh{\xi}z-\wb{\xi}z\big)+\wh{u}\big(\wb{u}\wh{\xi}-\wb{u}\wb{\xi}-\wt{\xi}z+\wb{\xi}z\big)\big]/ \nn \\
\hphantom{\wb{\wh{\wt{u}}}=}{}
/\big[\wt{u}\wh{u}\wt{z}-\wt{u}\wb{u}\wt{z}-\wh{u}z\wt{z}+\wb{u}z\wt{z}-\wt{u}\wh{u}\wh{z}
+\wh{u}\wb{u}\wh{z}+\wt{u}z\wh{z}-\wb{u}z\wh{z}+(\wt{u}-\wh{u})(\wb{u}-z)\wb{z} \big], \\
 \wb{\wh{\wt{z}}}=\frac{1}{\frac{\wb{\xi}-\wt{\xi}}{\wt{z}-\wb{z}}
+\frac{\wh{\xi}-\wb{\xi}}{\wh{z}-\wb{z}}} \bigg[G_4(-q,-a)/(\wh{u}-z)(\wh{z}-\wb{z})
-G_4(-p,-a)/(\wt{u}-z)(\wt{z}-\wb{z})\nn\\
\hphantom{\wb{\wh{\wt{z}}}=}{}
 +\frac{(\alpha_{3}-u)(\wb{u}-z)\big({-}\wb{\xi}\wt{z}
+\big(\wb{\xi}-\wt{\xi}\big)\wh{z}+\wh{\xi}(\wt{z}-\wb{z})+\wt{\xi}\wb{z}\big)
+(\wt{z}-\wh{z})G_4(-r,-a)}{(\wb{u}-z)(\wt{z}-\wb{z})(\wb{z}-\wh{z})}\bigg],\\
 \wb{\wh{\wt{v}}}= \frac{z \big((\wh{z}-\wt{z})\wb{u}\wb{\xi}
+\big(\wb{z}-\wh{\xi}\big)z\wt{z}+(\wt{z}-\wb{z})z\wh{z}+\big(\wt{\xi}-\wh{\xi}\big)(\wb{u}-z)\wb{z}\big)+C_{1}+C_{2}}
{(\wh{u}-\wb{u})\wt{u}\wt{z}+(\wb{u}-\wh{u})z\wt{z}+(\wb{u}-\wt{u})\wh{u}\wh{z}
+(\wt{u}-\wb{u})z\wh{z}+(\wt{u}-\wh{u})(\wb{u}-z)\wb{z}},
\end{gather*}
where
\begin{align*}
&C_{1}=\wt{u} \big[ \big(\wb{u}\wb{\xi}-\wh{u}\wh{\xi}\big)\wt{z}
+(z-\wb{u})\wt{\xi}\wb{z}-(z-\wh{u})\wt{\xi}\wh{z}+\big(\wt{\xi}-\wb{\xi}\big)z\wt{z}\big],\\
&C_{2}=\wh{u} \big[-\wb{u}\wb{\xi}\wh{z}+\big(-\wt{\xi}+\wb{\xi}\big)z\wh{z}
+\wh{\xi}(\wt{z}-\wb{z})z+\wb{u}\wb{z}\big].
\end{align*}
For the variable $\xi$, the formula for $\wb{\wh{\wt{\xi}}}$ is
somewhat long to be listed here.
The coincident relation $\wb{\wh{\wt{\xi}}}=\wh{\wb{\wt{\xi}}}=\wt{\wb{\wh{\xi}}}$
holds only if \eqref{eq:4comp-A-2 bc} and
its $(\wb{\phantom{a}}, \wh{~~})$ and $(\wb{\phantom{a}}, \wt{~~})$ versions hold,
i.e.,
\[ \wt{v}-\wh{v}=u(\wt{u}-\wh{u}),\qquad
\wt{v}-\wb{v}=u(\wt{u}-\wb{u}),\qquad
\wh{v}-\wb{v}=u(\wh{u}-\wb{u}). \]
This indicates that the four-component GD-4 (A-2) equation \eqref{eq:4comp-A-21} as a system is MDC;
\eqref{eq:4comp-A-2 bc} is not needed for determining the up-right evolution of~\eqref{eq:4comp-A-21},
but it is necessary for the whole system~\eqref{eq:4comp-A-21} to be MDC.

{\textbf{Alternative four-component GD-4 (A-2) equation \eqref{eq:4comp-A-22}:}}
For $z, w$ and $\xi$, their triple shifts are uniquely expressed as
\begin{gather*}
 \wb{\wh{\wt{z}}}=\frac{\wb{\varpi}(\wt{z}-\wh{z})(w-\wb{z})-\wh{\varpi}(w-\wh{z})(\wt{z}-\wb{z})
+\wt{\varpi}(w-\wt{z})(\wh{z}-\wb{z})}
{(\wt{z}-\wb{z})\wh{w}\wh{z}+(\wh{z}-\wt{z})\wb{w}\wb{z}+(\wb{z}-\wh{z})\wt{w}\wt{z}
+w \big (\wt{z}-\wb{z})\wb{w}+(\wh{z}-\wb{z})\wt{w}+(\wb{z}-\wt{z})\wh{w}\big)}, \\
 \wb{\wh{\wt{w}}}=\frac{1}{\frac{\wt{\varpi}-\wb{\varpi}}{\wt{w}-\wb{w}}
+\frac{\wb{\varpi}-\wh{\varpi}}{\wh{w}-\wb{w}}}
\bigg[ \frac{1}{\wb{w}-\wh{w}}\bigg (\frac{(\wt{w}-\wh{\varpi})\wb{\varpi}
+(\wh{w}-\wb{w})\wt{\varpi}+\wh{\varpi}(\wb{w}-\wt{w})(\alpha_{3}-z)}{\wt{w}-\wb{w}}
-\frac{G4(-q,-b)}{w-\wh{z}}\bigg) \nn\\
\hphantom{\wb{\wh{\wt{w}}}=}{}
 +\frac{1}{\wt{w}-\wb{w}}
\bigg(\frac{(\wh{w}-\wt{w})G_4(-r,-b)}{(\wh{w}-\wb{w})(w-\wb{z})}-\frac{G_4(-p,-b)}{w-\wt{z}}\bigg)\bigg],\\
 \wb{\wh{\wt{\xi}}}=\frac{-\wb{\varpi}(w \wt{w}-w\wh{w}-\wt{w}\wt{z}
+\wh{w}\wh{z})(w-\wb{z})+\wt{\varpi}(w-\wt{z})(-w \wh{w}+w\wb{w}+\wh{w}\wh{z}-\wb{w}\wb{z})+C_{3}}
{\wh{w}\wt{z}\wh{z}-\wb{w}\wt{z}\wb{z}-\wh{w}\wh{z}\wb{z}
+\wb{w}\wh{z}\wb{z}+\wt{w}\wt{z}(\wb{z}-\wh{z})+w\big(\wb{w}\wt{z}
+\wt{w}\wh{z}-\wb{w}\wh{z}-\wt{w}\wb{z}+\wh{w}(\wb{z}-\wt{z})\big)},
\end{gather*}
where
\begin{gather*}
 C_{3}= \wb{\varpi}(w-\wh{z})(w(\wt{w}-\wb{w})-\wt{w}\wt{z}+\wb{w}\wb{z}).
\end{gather*}
For the variable $\varpi$, the formula for $\wb{\wh{\wt{\varpi}}}$ is
long and one can check the coincident relation
$\wb{\wh{\wt{\varpi}}}=\wh{\wb{\wt{\varpi}}}=\wt{\wb{\wh{\varpi}}}$
holds only if \eqref{eq:4comp-3.41bb} and
its $(\wb{\phantom{a}}, \wh{~~})$ and $(\wb{\phantom{a}}, \wt{~~})$ versions hold,
i.e.,
\[ \wh{\xi}-\wt{\xi}=(\wh{z}-\wt{z})z, \qquad
\wb{\xi}-\wt{\xi}=(\wb{z}-\wt{z})z,\qquad
\wh{\xi}-\wb{\xi}=(\wh{z}-\wb{z})z. \]
In conclusion,
the four-component alternative GD-4 (A-2) equation \eqref{eq:4comp-A-22} as a system is MDC;
although \eqref{eq:4comp-3.41bb} is not needed for determining the up-right evolution of~\eqref{eq:4comp-A-22},
it is necessary for the whole system~\eqref{eq:4comp-A-22} to be MDC.

\subsection*{Acknowledgements}

The authors are grateful to the referees for their invaluable comments.
This work is supported by the National Natural Science Foundation of China (grant nos.~12271334,
12071432, 11875040).

\pdfbookmark[1]{References}{ref}
\LastPageEnding

\end{document}